%% file: main.tex
\documentclass[sigconf]{_acm/acmart}

\usepackage{bigstrut}
\usepackage{booktabs}
\usepackage{multirow}
\usepackage{microtype}
\usepackage{graphicx}
\usepackage{subfig}
\usepackage{bbding}
\usepackage{overpic}
\usepackage{balance}
\usepackage{amsthm}
\usepackage{array}
\usepackage{clrscode}
\usepackage{multicol}
\usepackage{float}
\usepackage{newfloat}
\usepackage{color}
\usepackage{xcolor}
\usepackage{amsopn}
\usepackage{mathrsfs}
\usepackage{mathtools}
\usepackage{amsmath}
\usepackage{arydshln}
\usepackage{blkarray}
\usepackage{enumerate}
\usepackage{courier}
\usepackage{rotating}
\usepackage{bm}
\usepackage{wrapfig}
\usepackage{ragged2e}
\usepackage[misc]{ifsym}
\usepackage{threeparttable}
\usepackage{makecell}
\usepackage{adjustbox} 
\usepackage{enumitem}
\usepackage{url}
\usepackage{xspace}
\usepackage{comment}
\usepackage{changepage}
\usepackage{algorithm}
\usepackage{algpseudocode} 
\usepackage{algorithmicx}

\usepackage{amssymb}
\usepackage{utfsym}
\usepackage[ruled, vlined, nofillcomment, linesnumbered, algo2e]{algorithm2e}
\usepackage[normalem]{ulem}
\usepackage{hyperref}

\useunder{\uline}{\ul}{}

\newcommand{\ie}{\emph{i.e.,}\xspace}
\newcommand{\eg}{\emph{e.g.,}\xspace}

\newcommand{\paratitle}[1]{\vspace{1.5ex}\noindent\textbf{#1}}

\newcommand{\ignore}[1]{}

\AtBeginDocument{%
  \providecommand\BibTeX{{%
    \normalfont B\kern-0.5em{\scshape i\kern-0.25em b}\kern-0.8em\TeX}}}

\copyrightyear{2018}
\acmYear{2018}
\setcopyright{acmcopyright}
\acmConference[Conference acronym 'XX]{Make sure to enter the correct conference title from your rights confirmation email}{June 03--05, 2018}{Woodstock, NY}
\acmBooktitle{Woodstock '18: ACM Symposium on Neural Gaze Detection, June 03--05, 2018, Woodstock, NY} 
\acmDOI{0000000.0000000}
\acmISBN{978-1-4503-XXXX-X/18/06}
\acmPrice{15.00}

\begin{document}

\title{LARES: Latent Reasoning for Sequential Recommendation}

\author{Enze Liu$^*$}
\orcid{0009-0007-8344-4780}
\affiliation{%
    \institution{GSAI, Renmin University of China}
    \city{Beijing}
    \country{China}
}
\email{enzeliu@ruc.edu.cn}

\author{Bowen Zheng$^*$}
\orcid{0009-0002-3010-7899}
\affiliation{%
    \institution{GSAI, Renmin University of China}
    \city{Beijing}
    \country{China}
}
\email{bwzheng0324@ruc.edu.cn}

\author{Xiaolei Wang$^{*}$}
\affiliation{%
    \institution{GSAI, Renmin University of China}
    \city{Beijing}
    \country{China}
}
\email{xiaoleiwang@ruc.edu.cn}

\author{Wayne Xin Zhao\textsuperscript{\Letter}}
\orcid{0000-0002-8333-6196}
\affiliation{
    \institution{GSAI, Renmin University of China}
    \city{Beijing}
    \country{China}
}
\email{batmanfly@gmail.com}

\author{Jinpeng Wang}
\affiliation{
    \institution{Meituan}
    \city{Beijing}
    \country{China}
}
\email{wangjinpeng04@meituan.com}

\author{Sheng Chen}
\affiliation{
    \institution{Meituan}
    \city{Beijing}
    \country{China}
}
\email{chensheng19@meituan.com}

\author{Ji-Rong Wen}
\orcid{0000-0002-9777-9676}
\affiliation{
    \institution{
    GSAI, Renmin University of China}
    \city{Beijing}
    \country{China}
}
\email{jrwen@ruc.edu.cn}

\thanks{$^*$ Equal contribution.}
\thanks{\Letter \ Corresponding author.}
\thanks{GSAI is the abbreviation of Gaoling School for Artificial Intelligence.}

\renewcommand{\shortauthors}{Enze Liu, et al.}

\begin{abstract}
Sequential recommender systems have become increasingly important in real-world applications that model user behavior sequences to predict their preferences. However, existing sequential recommendation methods predominantly rely on non-reasoning paradigms, which may limit the model's computational capacity and result in suboptimal recommendation performance. To address these limitations, we present \textbf{LARES}, a novel and scalable \underline{LA}tent \underline{RE}asoning framework for \underline{S}equential recommendation that enhances model's representation capabilities through increasing the computation density of parameters by depth-recurrent latent reasoning. Our proposed approach employs a recurrent architecture that allows flexible expansion of reasoning depth without increasing parameter complexity, thereby effectively capturing dynamic and intricate user interest patterns. A key difference of LARES lies in refining all input tokens at each implicit reasoning step to improve the computation utilization. To fully unlock the model's reasoning potential, we design a two-phase training strategy: (1) Self-supervised pre-training (SPT) with dual alignment objectives; (2) Reinforcement post-training (RPT). During the first phase, we introduce \textit{trajectory-level alignment} and \textit{step-level alignment} objectives, which enable the model to learn recommendation-oriented latent reasoning patterns without requiring supplementary annotated data. The subsequent phase utilizes reinforcement learning (RL) to harness the model's exploratory ability, further refining its reasoning capabilities. Comprehensive experiments on real-world benchmarks demonstrate our framework's superior performance. Notably, LARES exhibits seamless compatibility with existing advanced models, further improving their recommendation performance. Our code is available at \href{https://anonymous.4open.science/r/LARES-E458/}{\textcolor{blue}{https://anonymous.4open.science/r/LARES-E458/}}.
\end{abstract}




\maketitle

\input{sections/1-introduction.tex}
\input{sections/2-methodology.tex}

\input{sections/3-experiments.tex}
\input{sections/4-related-work.tex}
\input{sections/5-conclusion.tex}

\bibliographystyle{_acm/ACM-Reference-Format}
\bibliography{bibliography}


\end{document}

%% file: sections/1-introduction.tex
\section{Introduction}
\label{sec:introduction}

In the era of information explosion, recommender systems have emerged as indispensable components across real-world applications ranging from e-commerce platforms to online streaming services~\cite{e-commerce,lsvcr}. Among them, current research has increasingly focused on the analysis of evolving user behaviors for capturing latent intentions and sequential patterns to predict future interactions, which is termed as sequential recommendation. Recent years have witnessed significant advancements in this area, with notable methods like SASRec~\cite{sasrec} and BERT4Rec~\cite{bert4rec}, which adopt transformer-based architectures for improved user behavior modeling.

To cope with the intricate and dynamic nature of user behaviors, extensive research~\cite{hstu,astro,scale-law4sr} has been conducted to scale the computational capacity of sequential recommenders, thereby strengthening their representational power. Previous works~\cite{scale-law4sr,hstu,lum} have primarily pursued this goal through parameter scaling. However, these efforts have failed to replicate the success achieved in large language models (LLMs). This discrepancy primarily stems from unique challenges in recommendation systems, including inherent data sparsity and quality limitations that impede effective model scaling~\cite{scale-law4sr}.

Recent advances in large reasoning models~\cite{deepseek-r1,kimi-k1.5,still-3} have demonstrated that scaling test-time computation is another effective approach to increasing the utilization of computing power and can significantly enhance the reasoning capabilities of LLMs.
This suggests a promising approach to improving model performance by increasing the computation density for each parameter.
There are two primary paradigms for test-time scaling in LLMs: one is explicit reasoning~\cite{deepseek-coder,grpo,bot,reasonflux}, where models verbalize intermediate reasoning steps (\ie chain-of-thoughts (CoTs)) by generating meaningful tokens before producing final answers, and the other is latent reasoning~\cite{coconut,softcot,recurrent-reasoning}, where models perform multi-step implicit reasoning in the latent space without generating explicit reasoning tokens.
While most LLMs adopt explicit CoT reasoning, this approach faces challenges in recommendation systems.
Unlike LLMs operating in \textit{dense} textual spaces, most sequential recommenders are confined to \textit{sparse} item ID spaces. This fundamental difference makes it hard to define meaningful reasoning steps like CoTs and provide supervision signals for training~\cite{verify-step-by-step,omegaprm}.
Therefore, we adopt the latent reasoning approach to scaling sequential recommenders.
Recent work like ReaRec~\cite{rearec} has demonstrated the potential of this paradigm through autoregressive generation of implicit reasoning tokens.
Despite its effectiveness, this method remains computationally suboptimal as it only enriches \textit{one} token per reasoning step. 

To tackle this problem, we aim to fully unleash computation by leveraging \textit{all} input tokens at each reasoning step.
Our approach is inspired by the recently proposed depth-recurrent model architecture for latent reasoning~\cite{recurrent-reasoning}, which designs a recurrent block consisting of Transformer layers.
It repeatedly applies the recurrent block to update the hidden states of all the tokens in the latent space for test-time compute scaling.
For the scaling of sequential recommenders, since we also aim to unleash the computation of each item token in the latent space, it is feasible to develop a depth-recurrent architecture over existing recommendation models.

To this end, we propose \textbf{LARES}, a novel and scalable \textbf{LA}tent \textbf{RE}asoning approach for \textbf{S}equential recommendation that enables flexible test-time scaling by thinking in continuous latent spaces. 
Our approach adopts a recurrent architecture comprising two key components: a pre-block for initial processing and a core-block for iterative refinement. The core-block supports arbitrary iteration depth, allowing for dynamic computation scaling while improving computational density (\ie the amount of computation per parameter). To fully unlock the model’s reasoning capabilities, we develop a two-stage training strategy, the self-supervised pre-training (SPT) and reinforcement post-training (RPT). During the SPT stage, we propose the trajectory-level alignment and step-level alignment objectives to equip the model with recommendation-oriented latent reasoning patterns. Specifically, in trajectory-level alignment, we want to achieve knowledge transfer between high-quality reasoning processes. In step-level alignment, we aim to improve the thinking coherence among different intermediate steps. During the RPT stage, we employ reinforcement learning to further refine the model's reasoning capabilities for recommendation.

In summary, our work makes the following main contributions:

$\bullet$ We propose \textbf{LARES}, a novel scalable latent reasoning approach for sequential recommendation that leverage all the input tokens to perform multi-step reasoning in latent space with arbitrary depth.

$\bullet$ We design a two-stage training strategy including SPT and RPT to fully unleash the model's reasoning capabilities. During SPT, we introduce trajectory-level and step-level alignment to improve reasoning coherence. For RPT, we leverage RL to further improve recommendation performance via task-aligned rewards.

$\bullet$ Extensive experiments on real-world benchmarks validate LARES's superiority, demonstrating its effectiveness and seamless compatibility with existing sequential recommenders.

%% file: sections/2-methodology.tex
\begin{figure*}[]
    \centering
    \includegraphics[width=0.9\linewidth]{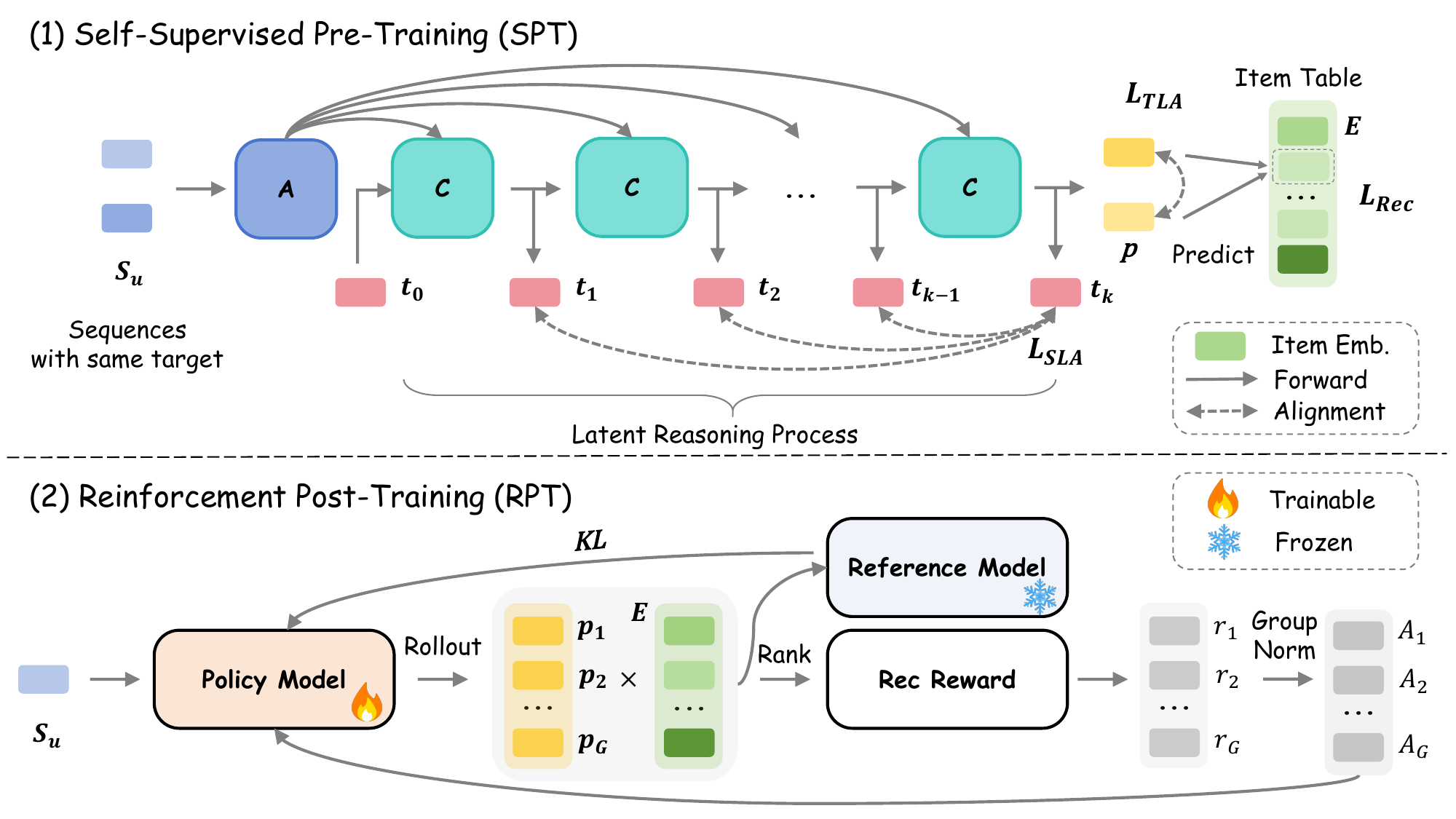}
    \captionsetup{font={small}}
    \caption{Overall framework of LARES. $\mathcal{A}$ and $\mathcal{C}$ denote the pre-block and core-block, repsectively. ``TLA'' and ``SLA'' represent the \textit{Trajectory-Level Alignment} and \textit{Step-Level Alignment}.}
    \label{fig:model}
\end{figure*}

\section{Methodology}
\label{sec:methodology}

In this section, we present our scalable latent reasoning approach for sequential recommendation, named \textbf{LARES}, which is illustrated in Figure~\ref{fig:model}.

\subsection{Overview}
\paratitle{Sequential Recommendation.} 
In recommender systems, there are a set of users $\mathcal{U}$ and a set of items ${\mathcal{V}}$. Let $M=|\mathcal{U}|$ and $N=|\mathcal{V}|$ denote the size of the user set and item set. The behavior record for each user $u\in\mathcal{U}$ is defined as an interaction sequence $S_u=[v_1,\dots,v_n]$ at the time step $n$, where items are arranged in chronological order.
As a typical sequential recommendation setting, traditional sequential recommenders~\cite{sasrec,s3rec,bert4rec} encode the interaction sequence by direct inference to obtain sequential behavior representations. The recommender then predicts the next item the user is most likely to interact with based on the similarities between the encoded user representations and candidate item representations. The objective of next item prediction can be formally written as:
\begin{equation}
    \max_{\Theta} P(v_{n+1}|S_u;\Theta),
\end{equation}
where $\Theta$ denotes the parameters of the sequential recommender.

\paratitle{Latent Reasoning For Sequential Recommendation.}
Traditional sequential recommenders typically adopt a straightforward inference pattern to directly provide recommendations, which struggle with complex recommendation tasks. In contrast, recent studies~\cite{rearec} propose a latent reasoning sequential recommender which is capable of thinking in a continuous space before making recommendations. This multi-step thinking paradigm allows the model to iteratively refine its inference in latent space and obtain more accurate user interests, simulating human-like mental thinking processes when addressing challenging problems. 
Generally, the model progressively derives a series of intermediate thoughts, denoted as $T_u$ and makes final recommendations conditioned on these latent thoughts as well as the behavior sequence. Formally, the objective of latent reasoning sequential recommenders can be formulated as:
\begin{equation}
    \max_{\Theta} P(v_{n+1}|S_u;\Theta)=P(v_{n+1}|T_u,S_u;\Theta)\cdot P(T_u|S_u;\Theta).
\end{equation}
Existing work~\cite{rearec} models latent reasoning $P(T_u|S_u;\Theta)$ as an autoregressive generation process, where a new state is generated and appended to the input at each reasoning step, which can be represented as follows:
\begin{equation}
    P(T_u|S_u;\Theta)=\prod_{i=1}^kP(t_i|S_u,t_{<i};\Theta),
\end{equation}
where $k$ is the number of reasoning steps and $t_i$ and $t_{<i}$ denote the generated thoughts at the $i$-th step and preceding the $i$-th step, respectively.

\paratitle{Scaling Latent Reasoning With Recurrent Depth.}
To fully unleash the computing power of deep thinking in sequential recommendation, we propose LARES, a novel latent reasoning scaling paradigm with recurrent depth where \textit{all} input tokens are refined at each reasoning step instead of only generating \textit{one} new token. The reasoning process of LARES is denoted as follows:
\begin{equation}
    P(T_u|S_u;\Theta) = \prod_{i=1}^k P(S^i_u|S^{i-1}_u, S^0_u;\Theta),
\end{equation}
where $k$ is the number of reasoning steps (\ie recurrent depth), $S^i_u$ is the thought at the $i$-th step, and $S^0_u$ corresponds to the initial input.
The overall framework of LARES is depicted in Figure~\ref{fig:model}. It adopts a depth-recurrent design consisting of a pre-block for mapping initial features into latent space and a core-block iterable to arbitrary depths, enabling flexible test-time scaling without additional parameters.
To facilitate effective latent reasoning with only the outcome label (\ie the next item), we propose a two-stage training approach for LARES: self-supervised pre-training for thinking adaptation and reinforcement post-training for thinking exploration.
This approach draws inspiration from established practices in LLMs, where pre-training enables LLMs to acquire fundamental knowledge, while reinforcement learning incentivizes the reasoning capability for complex reasoning tasks~\cite{openai-o1,deepseek-r1,kimi-k1.5,still-3}.
Accordingly, our framework first employs self-supervised pre-training to equip LARES with the core capability of user interest modeling through iterative latent reasoning, then applies reinforcement post-training to further incentivize its reasoning capability by exploring diverse latent thought patterns.

The remainder of this paper is organized as follows:
In Section~\ref{sec:arch}, we introduce the depth-recurrent architecture of the latent reasoning recommender. Section~\ref{sec:pretrain} presents the details of self-supervised pretraining, including \textit{trajectory-level alignment} and \textit{step-level alignment}, enabling the model to possess the basic ability to infer user interests through a multi-step latent thinking process. In Section~\ref{sec:rl}, we describe the process of reinforcement post-training, which leverages reinforcement learning algorithms to encourage the model to explore diverse thinking paths and further enhance the recommendation performance.

\subsection{Latent Reasoning with A Recurrent Architecture}
\label{sec:arch}
Our proposed framework, LARES, is a depth-recurrent sequential model consisting of multiple Transformer layers, which can be adapted to other advanced sequential recommendation architectures, \eg FMLPRec~\cite{fmlp-rec}, BSARec~\cite{bsarec}, and TedRec~\cite{tedrec}, as proved in the experiments in Section~\ref{sec:exp-backbone}.
In LARES, there are two blocks, \ie the pre-block $\mathcal{A}$ and the core-block $\mathcal{C}$.
The pre-block first transforms the item embeddings into the latent space, and then the core-block performs a multi-step reasoning on the latent representations in a recurrent pattern. Finally, the last latent thought representation is employed for subsequent item prediction.  
The most significant design of LARES lies in the recurrence of the core block, which allows the model to dive into a deeper thinking of user behaviors without introducing extra parameter burdens.
This design is inspired by the findings that the reasoning performance is up to a large effective depth but not necessarily many parameters~\cite{loop-power,recurrent-reasoning}.
The advantages of the recurrent architecture are twofold: on the one hand, it enhances the model's computational expressiveness with no extra parameter memory cost; on the other hand, it enables more flexible inference scaling by controlling the recurrent depth.

Suppose the historical behavior sequence of user $u$ is $S_u=[v_1,\dots,v_n]$. 
Given the item embedding table $\bm{E}\in\mathbb{R}^{N\times d}$, the sequence $S_u$ is first transformed into item embeddings $\bm{E}_u=[\bm{e}_{1};\dots;\bm{e}_{n}]$ by table lookup, where $d$ is the embedding dimension and $[;]$ denotes the concatenation operation.
First, $\bm{E}_u$ along with the position embeddings $\bm{E}_p\in\mathbb{R}^{n\times d}$ is fed into the pre-block $\mathcal{A}$ to produce the initial latent item representation $\bm{H}\in\mathbb{R}^{t\times d}$, which is written as:
\begin{equation}
    \bm{H}=\mathcal{A} \left( \bm{E}_u+\bm{E}_p \right).
\end{equation}
Then, the recurrent core-block $\mathcal{C}$ performs iterative latent reasoning, taking $\bm{H}$ and the latent thought of the previous step $\bm{T}_{i-1}$ as inputs, formulated as:
\begin{align}
    \bm{T}_{i}&=\mathcal{C}\left(\operatorname{LN}\left(f\left(\bm{T}_{i-1},\bm{H}\right)\right)\right),\ \ \text{for}\ i\in\{1,\dots,k\}, \\
    \bm{T}_0&\sim\mathcal{N}(0, \sigma_1^2I),
\end{align}
where $i$ and $k$ denote the $i$th step and the number of total reasoning steps, $f$ is the aggregation function \eg addition and concatenation, $\operatorname{LN}[\cdot]$ denotes the layer normalization operation and $\sigma_1$ is the standard deviation of the normal distribution for initializing the random state $\bm{T}_0$.
The input $\bm{H}$ to core-block $\mathcal{C}$ in every reasoning step functions as a residual connection to ensure stable gradient backpropagation. 
To allow flexible scaling of inference-time reasoning, during training, we sample the iteration count \( k \) per training step from a \textit{log-normal Poisson distribution}. For a target mean iteration count \( \bar{k}+1 \) (where \( \bar{k} \in \mathbb{N} \)) and variance \( \sigma_2 \), the sampling procedure is defined as follows:  
\begin{align}  
    \xi &\sim \mathcal{N}\left(\log(\bar{k}) - \frac{1}{2}\sigma_2^2, \sigma_2^2\right), \\  
    k &\sim \mathcal{P}(e^{\xi}) + 1,  
\end{align}  
where \( \mathcal{N} \) and \( \mathcal{P} \) denote the normal and Poisson distributions, respectively. In our experiments, we set \( \sigma_2 = 0.5 \).  
This distribution predominantly samples values below \( \bar{k} \), while retaining a heavy tail that occasionally yields significantly higher iteration counts. To mitigate excessive memory consumption, we truncate \( k \) such that \( k = \min(k, 3\bar{k}) \).  In inference, the recurrent depth is set to $\bar{k}+1$ for all test samples.

Finally, the final latent thought state corresponding to the last item $v_{n+1}$ is regarded as the final user preference representation $\bm{p}\in\mathbb{R}^d$ denoted as $\bm{p}=\bm{T}_k[-1]$.
The output $\bm{p}$ is then used to compute the recommended probabilities of candidate items $\hat{y}_i$, and we adopt the widely used cross-entropy loss as the recommendation objective, which is formulated as:
\begin{align}
    \mathcal{L}_{\operatorname{Rec}}=-\log\hat{y}_{n+1}=-\log\frac{\exp({\bm{p}\cdot\bm{e}^T_{n+1})}}{\sum_{i=1}^N \exp({\bm{p}\cdot\bm{e}^T_i})},
\end{align}
where $\bm{e}_{n+1}$ denotes the target item embedding of $v_{n+1}$ that user $u$ will interact with at the time step $n+1$.

\subsection{Self-Supervised Pre-Training for Thinking Adaptation}
\label{sec:pretrain}

In the self-supervised pre-training (SPT) stage, LARES learns to perform recommendation-oriented latent reasoning for user interest modeling. However, relying solely on the recommendation objective $\mathcal{L}_\text{Rec}$ is not enough to ensure effective training because it cannot provide sufficient signals for the intermediate latent thinking process of LARES. To alleviate this problem, we propose two self-supervised optimization objectives, \ie \textit{trajectory-level alignment} and \textit{step-level alignment}, to provide auxiliary supervision.

\subsubsection{Trajectory-Level Alignment}
To strengthen LARES's reasoning capability, we introduce trajectory-level alignment to leverage complementary strengths from different reasoning trajectories.
We define different trajectories from two aspects:
On the one hand, stochastic elements like random initialization and dropout naturally lead to varied reasoning paths across forward passes for the identical input sequences;
On the other hand, the trajectory of different sequences sharing the same target item can also be regarded as positive views~\cite{duorec}.
Specifically, we align the final outputs of independent reasoning trajectories between positive pairs.
However, our experiments reveal that aligning two reasoning outcomes with different step lengths adversely impacts model performance. We posit that this occurs because the inconsistency in reasoning steps causes a misalignment between long-chain and short-chain reasoning. Specifically, forcing short-chain reasoning to capture the richer and more complex patterns in longer reasoning processes is inherently challenging. As a result, the model tends to degenerate toward short-chain reasoning, ultimately compromising its effectiveness.
To address this, we ensure consistent reasoning steps within each positive pair.
Formally, given two positive sequences $S_u$ and $\hat{S}_u$ and a shared reasoning step $k$, LARES produces final preference representations $\bm{p}=\bm{T}_k[-1]$ and $\hat{\bm{p}}=\hat{\bm{T}}_k[-1]$. We achieve the alignment between them based on the InfoNCE loss. The trajectory-level alignment objective is formulated as:
\begin{align}  
    F(\bm{x}, \bm{y}^+, \mathcal{B}_{\bm{y}}) &= -\log \frac{\exp(s(\bm{x}, \bm{y}^+)/\tau)}{\sum_{\bm{y} \in \mathcal{B}_y} \exp(s(\bm{x}, \bm{y})/\tau)}, \\  
    \mathcal{L}_\text{TLA} &= \frac{1}{2} \left( F(\bm{p}, \hat{\bm{p}},\mathcal{R}_{\hat{\bm{p}}}) + F(\hat{\bm{p}}, \bm{p},\mathcal{R}_{\bm{p}}) \right),
\end{align}
where $F(\cdot,\cdot,\cdot)$ denotes the InfoNCE loss function, \( s(\cdot, \cdot) \) is a similarity metric (\eg cosine or dot product), $\mathcal{R}_{\bm{p}}$ and $\mathcal{R}_{\hat{\bm{p}}}$ represent sample sets containing both positive and negative instances, and \( \tau \) is a temperature coefficient controlling the distribution sharpness.

\subsubsection{Step-Level Alignment}

Given a historical interaction sequence \( S_u \) and a sampled iteration number \( k \), LARES generates a sequence of latent thought representations \( \bm{T}_u = [\bm{t}_1, \dots, \bm{t}_k] \).
Ideally, these latent thoughts are progressively refined to converge toward the true user preference distribution.
While the model is expected to produce increasingly accurate latent representations as reasoning advances, intermediate states may occasionally diverge from the desired trajectory, leading to suboptimal or counterproductive reasoning steps.  
To address this issue, we introduce \textit{step-level alignment} to enforce coherence between intermediate states and the final output. Specifically, we uniformly sample an intermediate step \( b \) from \( \{1, \dots, k-1\} \) and optimize the alignment between \( \bm{t}_b \) and \( \bm{t}_k \) using an InfoNCE loss:  
\begin{align}  
    b &\sim \operatorname{Unif}\{1, 2, \dots, k-1\}, \\  
    \mathcal{L}_\text{SLA} &= \frac{1}{2} \left( F(\bm{t}_b, \bm{t}_k, \mathcal{B}_k) + F(\bm{t}_k, \bm{t}_b, \mathcal{B}_b) \right),  
\end{align}
where \( \operatorname{Unif} \) denotes the uniform distribution and \( \mathcal{B}_i,i\in\{b,k\} \) contains the $i$-th step latent reasoning representations of all instances in the same batch.

The overall objective for self-supervised pre-training is written as:
\begin{equation}
    \mathcal{L}_\text{SPT}=\mathcal{L}_\text{Rec}+\alpha\mathcal{L}_\text{TLA}+\gamma\mathcal{L}_\text{SLA},
\end{equation}
where $\alpha$ and $\gamma$ denote hyper-parameters to balance the weights among different objectives during optimizations, respectively.

\subsection{Reinforcement Post-Training for Thinking Exploration}
\label{sec:rl}

After the self-supervised pre-training stage, the model acquires latent reasoning patterns for sequential recommendation tasks.
However, this stage suffers from limited thinking exploration due to the absence of supervisory signals that guide the model in distinguishing between ``good'' and ``not good'' reasoning steps. Consequently, the model's exploratory potential remains underutilized.
To mitigate this limitation, we introduce a reinforcement learning-based post-training approach that enhances the model's reasoning capabilities through learning from experiences of high-quality reasoning trajectories.
The subsequent sections introduce the reinforcement learning algorithm, reward design, and data selection strategy.

\subsubsection{Reinforcement Learning Algorithm}

To balance performance and computational cost, we employ the Group Relative Policy Optimization (GRPO) algorithm~\cite{grpo} during reinforcement post-training.
For each input (\ie user interaction sequences $S_u$ in our case), GRPO samples a group of rollouts from the old policy $\pi_{\theta_\text{old}}$.
The current policy \( \pi_{\theta} \) (\ie the base model) is then updated by maximizing a reward function and regularized by the KL divergence from a reference policy \( \pi_{\text{ref}} \) (\ie the initial pre-trained model).
The objective of GRPO is formulated as:
\begin{align}
    \mathcal{J}_\text{GRPO}(\theta)&=\mathbb{E}_{x\sim D,\{y_{i}\}_{i=1}^G\sim\pi_{\theta}} \nonumber\\
    &\left[\frac{1}{G}\sum_{i=1}^G
    \left(\min\left(P\cdot A_i, C\cdot A_i\right)-\beta\ \mathbb{D}_\text{KL}
    \left(\pi_\theta||\pi_{{\text{ref}}}
    \right)
    \right)
    \right],
    \label{eq:grpo}
\end{align}
\begin{equation}
    P=\frac{\pi_\theta(y_i|x)}{\pi_{\theta_\text{old}}(y_i|x)}, C=\operatorname{clip}\left(\frac{\pi_\theta(y_i|x)}{\pi_{\theta_\text{old}}(y_i|x)},1-\epsilon,1+\epsilon\right),
\end{equation}
where $A$ is the advantage value, $x$ is the input, $y_i$ is the response generated by LLMs, and $\pi_\theta(y|x)=\prod_{j=1}^{|y_i|}\pi_\theta(y_{i,j}|x, y_{i, <j})$ is the generation probability of the response $y_i$ under policy $\pi$.
However, directly applying Eq.~\eqref{eq:grpo} to our task is infeasible because it requires computing the joint probability $\pi(y_i|x)$ with discrete tokens, which are absent in our setting.
To resolve this issue, we reformulate it as the joint probability of recommending the target item at each reasoning step, denoted as $\pi(y_i|x)=\pi(v_{n+1}|S_u)=\prod_{j=1}^k\pi(v_{n+1}|S_u,t_{i,j})$ where $v_{n+1}$ is the target item, $S_u$ denotes the input user sequence, $t_{i,j}$ represents the latent thought representation at $j$-th step of the $i$-th rollout and $k$ is the total number of reasoning steps.

\subsubsection{Reward Design} 
To maintain strict alignment with recommendation objectives, we directly employ standard recommendation metrics (namely NDCG@\textit{k} and Recall@\textit{k}) as reward signals.
We take the recommendation results of the last reasoning step and the target label to calculate these metrics.
Specifically, for $G$ rollout trajectories, we first take the last latent thought states corresponding to the last item in the input sequence as the final user representations $\{\bm{p}_1,\dots, \bm{p}_{G}\}$ and then obtain the recommendation probability distribution for each rollout by computing the similarity between $\bm{p}_i$ and the item embedding table $\bm{E}$, denoted as $P_i=\operatorname{softmax}(\bm{p}_i\cdot\bm{E}^T)\in\mathbb{R}^{N},i\in\{1,\dots,G\}$. Then we can obtain the item ranking list based on the probability distribution for every rollout, denoted as $\operatorname{L}_i=\operatorname{argsort} P_i$. The reward function is formally written as:
\begin{equation}
    r_i = m(v_{n+1}, \operatorname{L}_i), \quad m \in \{\text{NDCG}@k, \text{Recall}@k\}.
\end{equation}
The advantage value $A_i$ is computed as the z-score normalized reward within each group:
\begin{equation}
    A_i=\frac{r_i-\operatorname{mean}(\{r_1,\dots,r_G\})}{\operatorname{std}(\{r_1,\dots,r_G\})}.
\end{equation}

\subsubsection{Data Selection}
Recent work demonstrates that the difficulty of data is important to the effectiveness of RL~\cite{kimi-k1.5}.
Since labels in sequential recommendation are very sparse, it is important that the data have a balanced difficulty for a limited number of rollouts.
Considering this, we propose a data selection strategy that filters out hard training samples.
Specifically, we exclude instances where the pre-trained model fails to rank the target item within the top 100 positions across three independent inference trials.

\subsection{Time Complexity}

As described in Section~\ref{sec:arch}, the LARES framework comprises two key components: a pre-block and a core-block. To illustrate the computational complexity, we adopt the transformer architecture as a representative example due to LARES's compatibility with diverse model designs. The primary computational overhead in each transformer layer stems from the multi-head self-attention, with complexities of $\mathcal{O}(N^2d + Nd^2)$, where $N$ represents the sequence length and $d$ denotes the model dimension.  
Let $L_1$ and $L_2$ denote the number of transformer layers in the pre-block and core-block, respectively. Consequently, their computational complexities can be expressed as $\mathcal{O}(L_1(N^2d + Nd^2))$ and $\mathcal{O}(L_2(N^2d + Nd^2))$. For a reasoning process involving $K$ iterative steps, the overall time complexity of LARES scales as $\mathcal{O}((L_1 + L_2K)(N^2d + Nd^2))$.

%% file: sections/3-experiments.tex
\section{Experiments}
\label{sec:experiments}
In this section, we conduct extensive experiments and analysis to empirically demonstrate the effectiveness of LARES.

\subsection{Experiment Setup}

\subsubsection{Dataset}
We assess our proposed method on four subsets derived from the latest Amazon 2023 review dataset: ``Musical Instruments'', ``Video Games'', ``Baby Products'', and ``Industrial \& Scientific''. Following prior studies~\cite{s3rec,fmlp-rec,tedrec}, we employ 5-core filtering to exclude inactive users and unpopular items with fewer than five interactions to ensure a robust evaluation. All user interactions are grouped by user ID and sorted chronologically.
We truncate behavior sequences to a maximum of 20 items per user. Table~\ref{tab:data_statistics} summarizes the key statistics of the preprocessed datasets.

\subsubsection{Baseline Models}
To conduct a comprehensive evaluation of LARES's performance, we compare it with multiple sequential recommendation baselines, including both reasoning-based and non-reasoning approaches:
\noindent (1) \emph{Non-Reasoning methods}:
{\textbf{GRU4Rec}}~\cite{gru4rec} employs GRUs to capture user behavior patterns.
{\textbf{SASRec}}~\cite{sasrec} is a transformer-based model utilizing unidirectional multi-head self-attention to encode user interaction sequences.
{\textbf{BERT4Rec}}~\cite{bert4rec} is a bidirectional self-attentive model that employs masked prediction for sequence modeling.
{\textbf{FMLP-Rec}}~\cite{fmlp-rec} replaces traditional self-attention with filter-enhanced MLPs to improve behavior modeling.
{\textbf{BSARec}}~\cite{bsarec} leverages Fourier transforms to capture both high- and low-frequency information in user behavior sequences.
{\textbf{CL4SRec}}~\cite{cl4srec} first introduces contrastive learning for sequential recommendation through three augmentation strategies: item masking, reordering, and cropping.
{\textbf{DuoRec}}~\cite{duorec} combines unsupervised model-level dropout augmentation with hard positive sample selection.
\noindent (2) \emph{Reasoning-based methods}:
{\textbf{ERL}}~\cite{rearec} enhances sequential recommendations by aggregating multi-step implicit reasoning states.
{\textbf{PRL}}~\cite{rearec} improves reasoning capabilities for sequential recommendation through contrastive learning with noise-disturbed positive views and temperature annealing.
{\textbf{PRL++}} extends PRL by incorporating DuoRec's sampling strategy.

\begin{table}[]
    \centering
    \captionsetup{font={small}}
    \caption{Statistics of the preprocessed datasets. Avg.L represents the average length of user interaction sequences.}
    \resizebox{\columnwidth}{!}{
    \begin{tabular}{lrrrrr}
    \toprule
     Dataset    &\#Users   &\#Items   &\#Actions &Avg.L &Sparsity  \\
     \midrule
     Instrument &57,439  &24,587  &511,836  &8.91   &99.964\%  \\
     Scientific &50,985  &25,848  &412,947  &8.10   &99.969\% \\
     Game &94,762  &25,612  &814,586  &8.60 &99.966\% \\
     Baby &150,777  &36,013  &1,241,083  &8.23  &99.977\% \\
     \bottomrule
    \end{tabular}}
    \label{tab:data_statistics}
\end{table}

\begin{table*}[]
\captionsetup{font={small}}
\caption{Performance comparison of different methods. The best and second-best results are indicated in bold and underlined font, respectively. ``*'' denotes that the improvements are statistically significant with $p < 0.01$ in a paired t-test setting.}
\label{tab:main_result}
\resizebox{0.98\textwidth}{!}{%
\renewcommand\arraystretch{0.88}
\setlength{\tabcolsep}{2mm}{
\begin{tabular}{llccccccccccc}
\toprule
Dataset & Metric & GRU4Rec & BERT4Rec & SASRec & FMLP-Rec & CL4SRec & DuoRec & BSARec & ERL & PRL & PRL++ & LARES \\ \midrule \midrule
\multirow{6}{*}{Instrument} & Recall@5 & 0.0318 & 0.0289 & 0.0346 & 0.0366 & 0.0354 & 0.0381 & 0.0363 & 0.0342 & 0.0345 & {\ul 0.0385} & \textbf{0.0411}$^*$ \\
 & Recall@10 & 0.0514 & 0.0463 & 0.0536 & 0.0575 & 0.0552 & {\ul 0.0598} & 0.0564 & 0.0546 & 0.0551 & 0.0587 & \textbf{0.0636}$^*$ \\
 & Recall@20 & 0.0774 & 0.0697 & 0.0798 & 0.0858 & 0.0831 & {\ul 0.0891} & 0.0841 & 0.0813 & 0.0834 & 0.0875 & \textbf{0.0934}$^*$ \\
 & NDCG@5 & 0.0207 & 0.0182 & 0.0216 & 0.0233 & 0.0227 & 0.0244 & 0.0231 & 0.0216 & 0.0222 & {\ul 0.0245} & \textbf{0.0263}$^*$ \\
 & NDCG@10 & 0.0271 & 0.0238 & 0.0277 & 0.0300 & 0.0291 & {\ul 0.0314} & 0.0295 & 0.0282 & 0.0288 & 0.0310 & \textbf{0.0336}$^*$ \\
 & NDCG@20 & 0.0336 & 0.0297 & 0.0343 & 0.0372 & 0.0361 & {\ul 0.0388} & 0.0365 & 0.0349 & 0.0359 & 0.0382 & \textbf{0.0410}$^*$ \\ \midrule
\multirow{6}{*}{Scientific} & Recall@5 & 0.0205 & 0.0183 & 0.0248 & 0.0250 & 0.0261 & {\ul 0.0280} & 0.0267 & 0.0245 & 0.0258 & 0.0279 & \textbf{0.0297}$^*$ \\
 & Recall@10 & 0.0340 & 0.0310 & 0.0385 & 0.0404 & 0.0406 & 0.0431 & 0.0421 & 0.0389 & 0.0405 & {\ul 0.0441} & \textbf{0.0464}$^*$ \\
 & Recall@20 & 0.0536 & 0.0478 & 0.0583 & 0.0608 & 0.0602 & 0.0650 & 0.0632 & 0.0584 & 0.0612 & {\ul 0.0661} & \textbf{0.0705}$^*$ \\
 & NDCG@5 & 0.0132 & 0.0116 & 0.0150 & 0.0157 & 0.0168 & {\ul 0.0178} & 0.0160 & 0.0151 & 0.0161 & 0.0176 & \textbf{0.0191}$^*$ \\
 & NDCG@10 & 0.0175 & 0.0157 & 0.0194 & 0.0206 & 0.0214 & 0.0226 & 0.0209 & 0.0198 & 0.0208 & {\ul 0.0228} & \textbf{0.0245}$^*$ \\
 & NDCG@20 & 0.0225 & 0.0199 & 0.0244 & 0.0258 & 0.0263 & 0.0281 & 0.0262 & 0.0247 & 0.0260 & {\ul 0.0283} & \textbf{0.0305}$^*$ \\ \midrule
\multirow{6}{*}{Game} & Recall@5 & 0.0504 & 0.0466 & 0.0578 & 0.0560 & 0.0555 & {\ul 0.0592} & 0.0572 & 0.0555 & 0.0545 & 0.0587 & \textbf{0.0616}$^*$ \\
 & Recall@10 & 0.0808 & 0.0731 & 0.0926 & 0.0922 & 0.0884 & {\ul 0.0932} & 0.0917 & 0.0888 & 0.0872 & 0.0925 & \textbf{0.0972}$^*$ \\
 & Recall@20 & 0.1236 & 0.1114 & 0.1392 & {\ul 0.1399} & 0.1337 & 0.1388 & 0.1381 & 0.1326 & 0.1332 & 0.1385 & \textbf{0.1444}$^*$ \\
 & NDCG@5 & 0.0321 & 0.0297 & 0.0334 & 0.0343 & 0.0347 & {\ul 0.0368} & 0.0355 & 0.0341 & 0.0345 & 0.0367 & \textbf{0.0386}$^*$ \\
 & NDCG@10 & 0.0419 & 0.0382 & 0.0446 & 0.0460 & 0.0452 & {\ul 0.0477} & 0.0466 & 0.0448 & 0.0450 & 0.0475 & \textbf{0.0500}$^*$ \\
 & NDCG@20 & 0.0527 & 0.0478 & 0.0563 & 0.0580 & 0.0566 & {\ul 0.0592} & 0.0583 & 0.0559 & 0.0566 & 0.0591 & \textbf{0.0619}$^*$ \\ \midrule
\multirow{6}{*}{Baby} & Recall@5 & 0.0204 & 0.0176 & 0.0229 & 0.0233 & 0.0231 & {\ul 0.0240} & 0.0232 & 0.0217 & 0.0220 & 0.0239 & \textbf{0.0250}$^*$ \\
 & Recall@10 & 0.0338 & 0.0292 & 0.0371 & 0.0378 & 0.0368 & {\ul 0.0383} & 0.0374 & 0.0353 & 0.0357 & 0.0382 & \textbf{0.0401}$^*$ \\
 & Recall@20 & 0.0548 & 0.0475 & 0.0580 & {\ul 0.0596} & 0.0576 & {\ul 0.0596} & 0.0586 & 0.0557 & 0.0571 & 0.0590 & \textbf{0.0625}$^*$ \\
 & NDCG@5 & 0.0131 & 0.0112 & 0.0140 & 0.0146 & 0.0147 & {\ul 0.0153} & 0.0148 & 0.0135 & 0.0139 & 0.0146 & \textbf{0.0160}$^*$ \\
 & NDCG@10 & 0.0174 & 0.0149 & 0.0186 & 0.0192 & 0.0190 & {\ul 0.0199} & 0.0193 & 0.0178 & 0.0183 & 0.0192 & \textbf{0.0208}$^*$ \\
 & NDCG@20 & 0.0226 & 0.0195 & 0.0238 & 0.0247 & 0.0243 & {\ul 0.0252} & 0.0247 & 0.0230 & 0.0237 & 0.0245 & \textbf{0.0264}$^*$ \\
 \bottomrule 
\end{tabular}%
}
}
\end{table*}

\subsubsection{Evaluation Settings}

We evaluate the sequential recommendation task using two standard metrics: Recall@$K$ and NDCG@$K$, with $K\in\{5,10,20\}$. Following prior work~\cite{s3rec,tiger,cocorec}, we adopt a \emph{leave-one-out} strategy. Specifically, for each user interaction sequence, we use the most recent interaction as the test instance, the second-last interaction for validation, and all remaining historical interactions for training.
To ensure rigorous evaluation and mitigate potential biases from negative sampling, we perform full ranking over the entire item pool. The final results report the average metric scores across all test instances.

\begin{table*}[]
\centering
\captionsetup{font={small}}
\caption{Ablation studies of LARES on three datasets by selectively removing the Trajectory-Level Alignment (TLA), Step-Level Alignment (SLA), and Reinforcement Post-Training (RPT). ``N@K'' and ``R@K'' denote “NDCG@K” and “Recall@K”, respectively.}
\label{tab:ablation}
\resizebox{0.95\textwidth}{!}{%
\renewcommand\arraystretch{0.9}
\setlength{\tabcolsep}{1.8mm}{
\begin{tabular}{ccccccccccccccc}
\toprule
\multicolumn{3}{c}{Variants} & \multicolumn{4}{c}{Instrument} & \multicolumn{4}{c}{Scientific} & \multicolumn{4}{c}{Game} \\
\cmidrule(l){1-3} \cmidrule(l){4-7} \cmidrule(l){8-11} \cmidrule(l){12-15}
TLA & SLA & RPT & R@5 & R@10 & N@5 & N@10 & R@5 & R@10 & N@5 & N@10 & R@5 & R@10 & N@5 & N@10 \\ \midrule
 &  &  & 0.0356 & 0.0571 & 0.0221 & 0.0290 & 0.0245 & 0.0395 & 0.0148 & 0.0196 & 0.0559 & 0.0903 & 0.0337 & 0.0448 \\
\usym{2713} &  &  & 0.0379 & 0.0596 & 0.0242 & 0.0312 & 0.0253 & 0.0403 & 0.0157 & 0.0205 & 0.0600 & 0.0943 & 0.0374 & 0.0488 \\
\usym{2713} & \usym{2713} &  & {\ul 0.0396} & {\ul 0.0624} & {\ul 0.0252} & {\ul 0.0326} & {\ul 0.0290} & {\ul 0.0456} & {\ul 0.0180} & {\ul 0.0234} & {\ul 0.0604} & {\ul 0.0961} & {\ul 0.0380} & {\ul 0.0491} \\
\usym{2713} & \usym{2713} & \usym{2713} & \textbf{0.0411} & \textbf{0.0636} & \textbf{0.0263} & \textbf{0.0336} & \textbf{0.0297} & \textbf{0.0464} & \textbf{0.0191} & \textbf{0.0245} & \textbf{0.0616} & \textbf{0.0972} & \textbf{0.0386} & \textbf{0.0500} \\ \bottomrule
\end{tabular}}}
\end{table*}

\subsubsection{Implementation}

We implement LARES and all baseline models using PyTorch.
To ensure a fair comparison, the batch size, embedding size and hidden size are set to 1024, 64 and 256, respectively.
All models are optimized using the AdamW optimizer with a learning rate of 0.001. For self-attentive models, the number of attention heads is fixed at 2. For LARES, the number of layers for both the pre-block and core-block is selected from \(\{1, 2\}\), while the mean reasoning step \(\bar{k}\) is chosen from \(\{3, 4, 6\}\). The hyperparameter for latent reasoning state initialization \(\sigma_1\) is set to 1.
In SPT, we use a learning rate of 0.001 and a dropout rate of 0.5. The coefficients for trajectory-level alignment \(\alpha\) and step-level alignment \(\gamma\) are tuned within \(\{0.1, 0.2, 0.3\}\) and \(\{0.1, 0.3, 0.5, 0.7\}\), respectively. In RPT, the learning rate is selected from \(\{0.0005, 0.0003, 0.0001\}\), and \(\beta\) is tuned from \(\{0.5, 1.0\}\). The rollout number \(G\) is fixed at 4. The reward function is selected from $\{\operatorname{Recall@5}, \operatorname{Recall@10}\}$. 
The non-reasoning SR baselines are implemented using RecBole~\cite{recbole,recbole2.0}, an open-source recommendation library, while reasoning-based SR models are reproduced from their official source codes. To mitigate overfitting, we employ early stopping, terminating training if NDCG@10 on the validation set shows no improvement for 10 consecutive epochs.

\subsection{Overall Performance}
We evaluate the performance of our proposed LARES framework by comparing it with various baseline methods across four real-world benchmark datasets. The comprehensive experimental results are presented in Table~\ref{tab:main_result}, from which we draw the following key observations:

\begin{itemize}[leftmargin=*]

\item For non-reasoning models, SASRec surpasses BERT4Rec and GRU4Rec across all datasets, confirming the superiority of transformer architectures in modeling user behavior sequences. And BERT4Rec underperforms both GRU4Rec and SASRec, suggesting that bidirectional self-attention mechanisms incorporating future context may be suboptimal for sequential recommendation tasks.
The filter-enhanced models (\ie FMLP-Rec and BSARec) perform better than SASRec, demonstrating the effectiveness of noise reduction in the frequency domain for capturing sequential patterns.
Contrastive learning approaches (\ie CL4SRec and DuoRec) consistently outperform traditional ID-based baselines, showing that contrastive regularization can enhance the SR model's representation learning. DuoRec's superior performance over CL4SRec stems from its robust model-level dropout augmentation strategy compared to sequence-level augmentations.

\item For reasoning models, ERL achieves superior performance than SASRec across most datasets, indicating that performing multi-step reasoning can better capture the user preference.
PRL consistently outperforms ERL demonstrating that noise-disturbed reasoning outcomes as contrastive signals can enhance the model's ability to extract critical sequence information.
PRL++ shows significant improvements over PRL, proving that incorporating hard positive samples effectively strengthens the effectiveness of contrastive learning.

\item Our proposed framework LARES outperforms all baselines, including both non-reasoning and reasoning models, by a large margin across all evaluation metrics on four datasets.
This consistent superiority underscores its advanced reasoning capabilities for sequential recommendation tasks.
LARES enables multi-step latent reasoning through core-block iteration at an arbitrary depth, allowing flexible computation scaling by increasing the computational density of parameters.
To fully exploit its latent reasoning potential, we design two training stages: self-supervised pre-training to instill latent reasoning patterns tailored for sequential recommendation tasks and reinforcement post-training to further stimulate its reasoning capabilities by encouraging thinking exploration.
These results also validate the effectiveness of depth-recurrent latent reasoning for sequential recommendation tasks.

\end{itemize}

\subsection{Ablation Studies}
We conduct an ablation study to evaluate the contribution of each key component in LARES: Trajectory-Level Alignment (TLA), Step-Level Alignment (SLA), and Reinforcement Post-Training (RPT). As shown in Table~\ref{tab:ablation}, we gradually remove these components and have the following findings:

\begin{itemize}[leftmargin=*]

\item The removal of any single component (TLA, SLA, or RPT) consistently degrades performance across all datasets. The complete removal of all components (leaving only the backbone model with latent reasoning) yields the worst performance, demonstrating that both alignment objectives in the pre-training stage are essential and RPT provides additional improvements.

\item The variant incorporating both TLA and SLA outperforms the version with TLA alone, confirming that these alignment strategies provide complementary benefits:
TLA enhances trajectory-level reasoning consistency and SLA improves step-level reasoning coherence.

\item The relative importance of TLA and SLA varies across domains. Specifically, TLA shows greater impact on Instrument and Game datasets, suggesting hard positive samples are particularly informative in these domains.
SLA proves more effective for Scientific dataset, indicating that reasoning coherence is especially valuable in Scientific scenario.

\end{itemize}

\begin{figure}[]
    \centering
    \includegraphics[width=\linewidth]{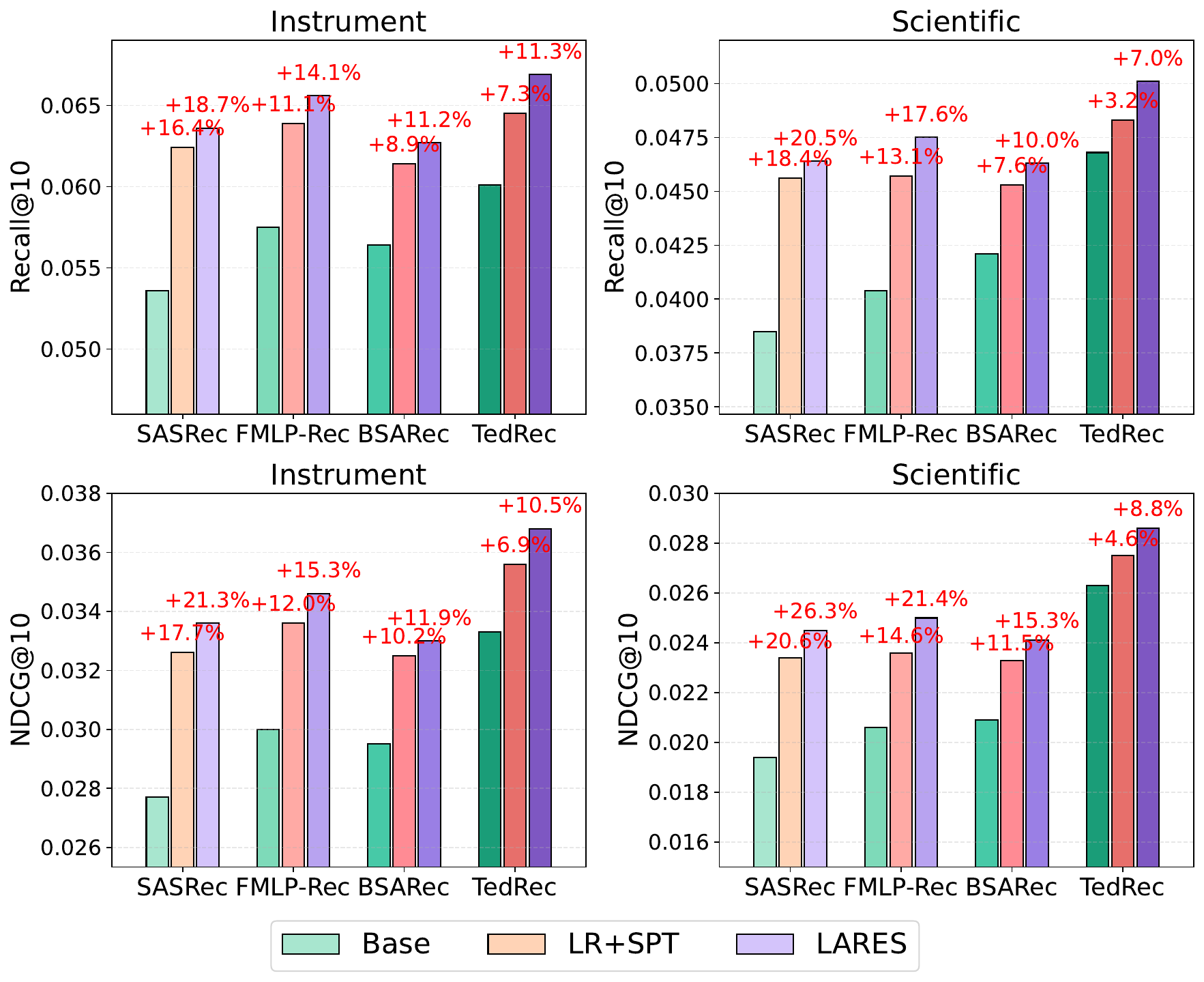}
    \caption{Performance comparison of LARES across different backbone architectures on Instrument and Scientific. `Base' indicates the original backbone model. `LR+SPT' refers to the backbone enhanced with our proposed latent reasoning module and pretraining. `LARES' denotes the full implementation incorporating all proposed components.}
    \captionsetup{font={small}}
    \label{fig:backbone}
\end{figure}

\subsection{Further Analysis}

\subsubsection{Compatibility with Different Backbones}
\label{sec:exp-backbone}
To evaluate the compatibility of LARES with different architectures, we conduct experiments using three advanced backbones in addition to SASRec (namely FMLP-Rec, BSARec, and TedRec~\cite{tedrec}).
The first two are conventional ID-based sequential recommendation models, whereas TedRec is a text-enhanced model that incorporates both ID embeddings and text embeddings derived from item metadata using a pretrained language model (\ie Sentence-T5 in our case).
Figure~\ref{fig:backbone} presents the performance comparison of three model variants: (1) the backbone (Base), (2) the backbone enhanced with latent reasoning through pretraining (LR+SPT), and (3) the complete framework (LARES).
The experimental results demonstrate that LARES consistently improves all four backbones, achieving performance gains of 15.3\% and 21.4\% in NDCG@10 for FMLP-Rec on the Instrument and Scientific datasets, respectively.
For BSARec, LARES yields over 10\% improvement across all metrics on both datasets. Notably, even the superior TedRec model benefits from our framework, showing average improvements of 10.9\% and 7.9\% on the respective datasets. These results substantiate the great compatibility of our latent reasoning paradigm.
Furthermore, we observe a consistent performance ranking across all metrics and datasets: Base < LR+SPT < LARES. This hierarchy validates the effectiveness of our proposed two-stage training strategy.

\begin{figure}[]
    \centering
    \includegraphics[width=\linewidth]{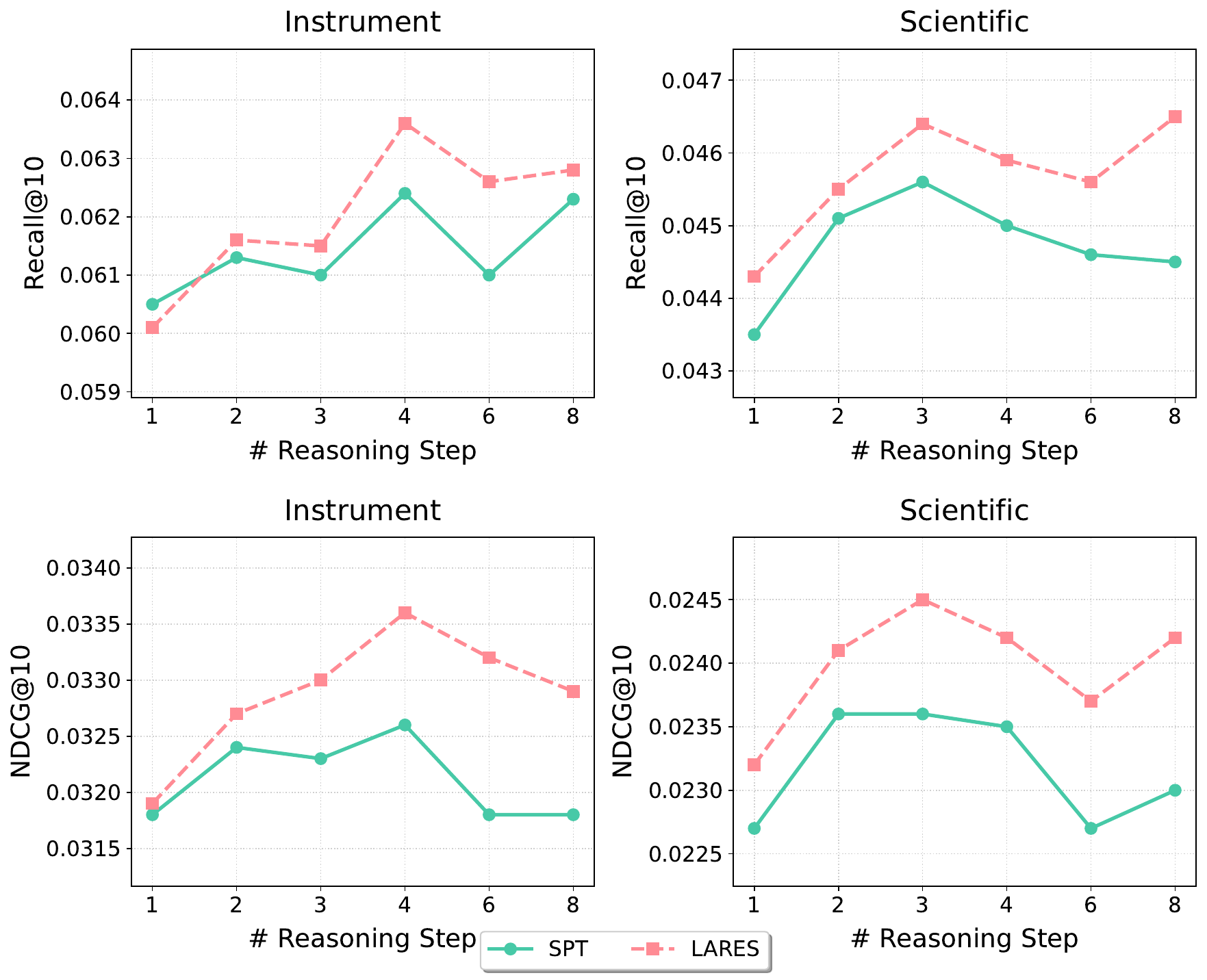}
    \caption{Performance of different reasoning steps on Instrument and Scientific. `SPT' denotes the latent reasoning model with only pre-training. `LARES' denotes the model with both pre-training and post-training.}
    \captionsetup{font={small}}
    \label{fig:step}
\end{figure}

\begin{figure}[]
    \centering
    \includegraphics[width=\linewidth]{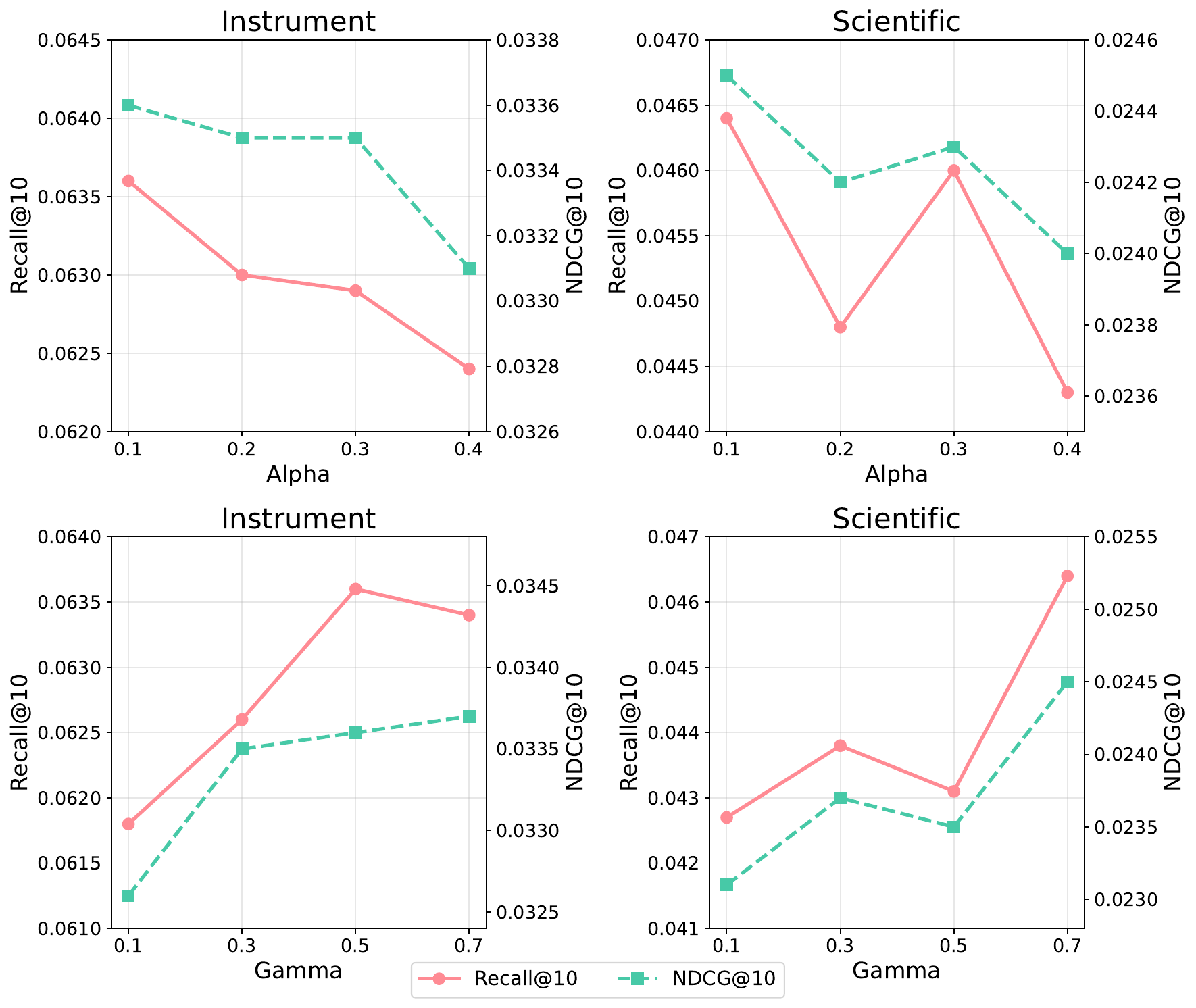}
    \caption{Performance of different alignment coefficients $\alpha$ and $\gamma$ on Instrument and Scientific.}
    \captionsetup{font={small}}
    \label{fig:coeff}
\end{figure}

\subsubsection{Influence of Reasoning Steps on Recommendation Performance}

To analyze the influence of reasoning step, we vary the number of reasoning steps \(\bar{k}\) in \(\{1, 2, 3, 4, 6, 8\}\) and evaluate the model on the Instrument and Scientific datasets. The results in Figure~\ref{fig:step} reveal a consistent trend: performance initially improves with more reasoning steps but declines after reaching an optimal point, which is in line with the findings in ReaRec.  
This pattern suggests that moderate reasoning step enhances model performance by strengthening representation power through additional computation. However, excessively large \(\bar{k}\) leads to performance degradation, likely because simple user interaction sequences do not require intensive reasoning-a phenomenon analogous to ``overthinking'' in NLP~\cite{overthinking}. The optimal reasoning steps are 4 for Instrument and 3 for Scientific, highlighting the importance of selecting an appropriate reasoning depth for optimal performance.  

\subsubsection{Influence of Alignment Coefficients}

We examine the effects of trajectory-level alignment and step-level alignment coefficients by evaluating model performance on Instrument and Scientific datasets across varying values $\alpha \in$ \{0.1, 0.2, 0.3, 0.4\} and $\gamma \in$ \{0.1, 0.3, 0.5, 0.7\}. 
From the results in Figure~\ref{fig:coeff}, we can observe that model performance on both datasets fluctuates with increasing $\alpha$ values, peaking at 0.1 and reaching the lowest point at 0.4. This suggests that excessive trajectory-level alignment may interfere with sequential pattern learning. 
For $\gamma$, the performance on Instrument initially improves and then declines at higher $\gamma$ values. The performance on Scientific exhibits generally consistent improvement with increasing $\gamma$ values.
These findings highlight the critical role of intra-step reasoning coherence in maintaining reasoning quality, while revealing differential sensitivity to alignment coefficients across datasets.

\subsubsection{Influence of Reasoning Steps on Inference Time}

To systematically assess the impact of reasoning steps on inference efficiency, we conduct a comparative analysis between LARES with varying reasoning steps and SASRec under equivalent computational budgets (FLOPs) on the Instrument dataset. 
LARES uses a 2-layer pre-block and core-block, and the batch size is set to 1024 for both.
As presented in Table~\ref{tab:time} (where item embeddings are excluded from parameter calculations), LARES achieves higher computational complexity while maintaining identical parameter sizes through increased reasoning steps.
For instance, with 4 reasoning steps, LARES attains an effective depth of 2+2×4 layers, comparable to a 10-layer SASRec architecture. The additional inference latency introduced by LARES remains moderate—approximately 10\% higher than SASRec at comparable FLOPs.
This computational overhead is acceptable in consideration of LARES's substantial performance gains, which improve SASRec's baseline by an average of 20\% on Instrument as shown in Figure~\ref{fig:backbone}.
Our findings demonstrate that LARES offers an advantageous trade-off between computational cost and model performance, suggesting strong potential for practical deployment.

\begin{table}[]
\captionsetup{font={small}}
\caption{Comparison of inference time between LARES with different reasoning steps and SASRec with the same FLOPs on Instrument. The experiments are conducted on a single RTX3090 GPU.}
\label{tab:time}
\begin{tabular}{lrrrc}
\toprule
Method & Step & Params & FLOPs & Inference Time \\ \midrule \midrule
\multirow{4}{*}{SASRec} & 1 & 4x & 4x & 1.10s \\
 & 1 & 6x & 6x & 1.23s \\
 & 1 & 8x & 8x & 1.40s \\
 & 1 & 10x & 10x & 1.55s \\ \midrule
\multirow{4}{*}{LARES} & 1 & 2x+2x & 4x & 1.19s \\
 & 2 & 2x+2x & 6x & 1.39s \\
 & 3 & 2x+2x & 8x & 1.58s \\
 & 4 & 2x+2x & 10x & 1.74s \\
 \bottomrule
\end{tabular}
\end{table}

%% file: sections/4-related-work.tex
\section{Related Work}
\label{sec:related}

\subsection{Sequential Recommendation}

Sequential recommendation has become a prominent research area in recommender systems, with the objective of modeling latent patterns in user behavior sequences to predict the next item of interest. Early approaches~\cite{fpmc, fossil} modeled user behaviors as Markov chains, focusing exclusively on item transition patterns. The advent of deep learning revolutionized this field, leading to the adoption of various neural architectures. These include convolutional neural networks (CNNs)~\cite{caser, cosrec,nextitnet}, recurrent neural networks (RNNs)~\cite{gru4rec, DBLP:conf/recsys/TanXL16, DBLP:conf/recsys/HidasiQKT16,hrnn}, and graph neural networks (GNNs)~\cite{DBLP:conf/sigir/ChangGZHNSJ021, DBLP:conf/aaai/WuT0WXT19}. Recently, Transformer-based models \cite{sasrec, bert4rec, DBLP:journals/tkde/HaoZZLSXLZ23} have demonstrated superior performance in sequential behavior modeling. Several recent studies~\cite{rec-denoiser,convformer,ac-tsr,fame} have proposed to enhance Transformer architectures. FMLP-Rec~\cite{fmlp-rec} replaces self-attention with filter-enhanced MLPs to reduce noise in user preference modeling.
However, these ID-based methods often suffer from cold-start problems. To address this, alternative approaches incorporate item textual metadata to enrich representations~\cite{unisrec,tedrec,cocorec}. 
UniSRec employs multi-domain textual data with MoE-enhanced adapters to learn universal sequential representations.
TedRec achieves sequence-level fusion of textual and ID representations through contextual convolution.
Despite these advancements, the test-time scaling in sequential recommendation is underexplored. In this paper, we propose a new latent reasoning paradigm for sequential recommendation that leverages all input tokens to perform multi-step reasoning in latent space with arbitrary depth.

\subsection{Reasoning Models}

Recent advances in test-time scaling~\cite{deepseek-r1,still-3,r1-like} have shifted the research focus in GenAI from large language models (LLMs) to large reasoning models (LRMs). LRMs excel in complex reasoning tasks through deep thinking capabilities, as demonstrated by the powerful reasoning systems like OpenAI-o1 and DeepSeek-R1. These models employ explicit long Chain-of-Thought mechanisms to generate extensive reasoning tokens before producing final answers~\cite{bot,got,reasonflux}. However, their reliance on explicit reasoning poses some critical challenges, including excessive memory demands from long context windows and limited expressive power due to discrete language space constraints.
To address these limitations, recent studies~\cite{coconut,latro,recurrent-reasoning,softcot} have introduced latent reasoning models that perform implicit reasoning in continuous latent spaces, achieving greater efficiency. In recommender systems, some efforts have adapted reasoning mechanisms for sequential recommendation models~\cite{rearec,stream-rec}. For instance, ReaRec~\cite{rearec} autoregressively generates latent reasoning tokens to refine user representations. STREAM-Rec~\cite{stream-rec} integrates slow-thinking paradigms with TIGER-style generative recommenders by producing annotated reasoning tokens before final recommendation semantic tokens.
In contrast, we propose \textbf{LARES}, a novel recurrent-depth latent reasoning framework for sequential recommendation. Unlike prior work, LARES iteratively refines all input tokens at each reasoning step. Additionally, it is seamlessly compatible with existing sequential recommendation models, further enhancing their performance.

%% file: sections/5-conclusion.tex
\section{Conclusion}
\label{sec:conclusion}

In this work, we propose \textbf{LARES}, a scalable and novel latent reasoning framework for sequential recommendations. A significant difference between LARES and previous latent reasoning methods like ReaRec lies in that our method leverages all input tokens during the latent reasoning process instead of generating only one reasoning token for higher computation utilization. Our approach adopts a depth-recurrent architecture consisting of a pre-block for mapping initial features into the latent space and an iterable core-block, enabling flexible test-time computation scaling without extra parameters. To fully exploit the potential of latent reasoning, we design a two-stage training pipeline including self-supervised pre-training (SPT) and reinforcement post-training (RPT). In the SPT stage, we introduce the trajectory-level alignment and step-level alignment to improve the model's reasoning capabilities from two different aspects. Trajectory-level alignment aims to align the reasoning outcomes of the user interaction sequences with the same target for mutual enhancement, while the target of step-level alignment is to improve the intra-step reasoning coherence to avoid divergences of intermediate reasoning steps. In the RPT stage, we leverage reinforcement learning to further stimulate model's reasoning abilities by encouraging it to explore diverse reasoning paths. To fully align with the downstream tasks, we directly adopt the standard metrics as rewards. Extensive experiments have demonstrated the superior performance of LARES and its seamless compatibility with existing advanced SR models.

Notably, there remain some promising directions for future research:
\begin{itemize}[leftmargin=*]
    \item \textbf{Adaptive Reasoning Depth.} While our current approach employs a fixed reasoning step for all user interaction sequences, we observe that not all sequences necessitate deep reasoning. Excessive reasoning steps may incur unnecessary computational overhead and even degrade recommendation performance. This raises an open question: \textit{How to adaptively determine the reasoning depth for different user behavior sequences?}
    \item \textbf{Process Reinforcement Learning.} In our RPT stage, we only consider the outcome rewards which are coarse-grained for the intermediate reasoning states. Therefore, incorporating process reinforcement learning could provide more precise guidance and improve the overall reasoning quality.
\end{itemize}

%% file: main.bbl

\begin{thebibliography}{55}


\ifx \showCODEN    \undefined \def \showCODEN     #1{\unskip}     \fi
\ifx \showISBNx    \undefined \def \showISBNx     #1{\unskip}     \fi
\ifx \showISBNxiii \undefined \def \showISBNxiii  #1{\unskip}     \fi
\ifx \showISSN     \undefined \def \showISSN      #1{\unskip}     \fi
\ifx \showLCCN     \undefined \def \showLCCN      #1{\unskip}     \fi
\ifx \shownote     \undefined \def \shownote      #1{#1}          \fi
\ifx \showarticletitle \undefined \def \showarticletitle #1{#1}   \fi
\ifx \showURL      \undefined \def \showURL       {\relax}        \fi
\providecommand\bibfield[2]{#2}
\providecommand\bibinfo[2]{#2}
\providecommand\natexlab[1]{#1}
\providecommand\showeprint[2][]{arXiv:#2}

\bibitem[Besta et~al\mbox{.}(2024)]%
        {got}
\bibfield{author}{\bibinfo{person}{Maciej Besta}, \bibinfo{person}{Nils Blach}, \bibinfo{person}{Ales Kubicek}, \bibinfo{person}{Robert Gerstenberger}, \bibinfo{person}{Michal Podstawski}, \bibinfo{person}{Lukas Gianinazzi}, \bibinfo{person}{Joanna Gajda}, \bibinfo{person}{Tomasz Lehmann}, \bibinfo{person}{Hubert Niewiadomski}, \bibinfo{person}{Piotr Nyczyk}, {and} \bibinfo{person}{Torsten Hoefler}.} \bibinfo{year}{2024}\natexlab{}.
\newblock \showarticletitle{Graph of Thoughts: Solving Elaborate Problems with Large Language Models}. In \bibinfo{booktitle}{\emph{Thirty-Eighth {AAAI} Conference on Artificial Intelligence, {AAAI} 2024, Thirty-Sixth Conference on Innovative Applications of Artificial Intelligence, {IAAI} 2024, Fourteenth Symposium on Educational Advances in Artificial Intelligence, {EAAI} 2014, February 20-27, 2024, Vancouver, Canada}}, \bibfield{editor}{\bibinfo{person}{Michael~J. Wooldridge}, \bibinfo{person}{Jennifer~G. Dy}, {and} \bibinfo{person}{Sriraam Natarajan}} (Eds.). \bibinfo{publisher}{{AAAI} Press}, \bibinfo{pages}{17682--17690}.
\newblock
\href{https://doi.org/10.1609/AAAI.V38I16.29720}{doi:\nolinkurl{10.1609/AAAI.V38I16.29720}}


\bibitem[Chang et~al\mbox{.}(2021)]%
        {DBLP:conf/sigir/ChangGZHNSJ021}
\bibfield{author}{\bibinfo{person}{Jianxin Chang}, \bibinfo{person}{Chen Gao}, \bibinfo{person}{Yu Zheng}, \bibinfo{person}{Yiqun Hui}, \bibinfo{person}{Yanan Niu}, \bibinfo{person}{Yang Song}, \bibinfo{person}{Depeng Jin}, {and} \bibinfo{person}{Yong Li}.} \bibinfo{year}{2021}\natexlab{}.
\newblock \showarticletitle{Sequential Recommendation with Graph Neural Networks}. In \bibinfo{booktitle}{\emph{{SIGIR} '21: The 44th International {ACM} {SIGIR} Conference on Research and Development in Information Retrieval, Virtual Event, Canada, July 11-15, 2021}}, \bibfield{editor}{\bibinfo{person}{Fernando Diaz}, \bibinfo{person}{Chirag Shah}, \bibinfo{person}{Torsten Suel}, \bibinfo{person}{Pablo Castells}, \bibinfo{person}{Rosie Jones}, {and} \bibinfo{person}{Tetsuya Sakai}} (Eds.). \bibinfo{publisher}{{ACM}}, \bibinfo{pages}{378--387}.
\newblock
\href{https://doi.org/10.1145/3404835.3462968}{doi:\nolinkurl{10.1145/3404835.3462968}}


\bibitem[Chen et~al\mbox{.}(2024)]%
        {latro}
\bibfield{author}{\bibinfo{person}{Haolin Chen}, \bibinfo{person}{Yihao Feng}, \bibinfo{person}{Zuxin Liu}, \bibinfo{person}{Weiran Yao}, \bibinfo{person}{Akshara Prabhakar}, \bibinfo{person}{Shelby Heinecke}, \bibinfo{person}{Ricky Ho}, \bibinfo{person}{Phil Mui}, \bibinfo{person}{Silvio Savarese}, \bibinfo{person}{Caiming Xiong}, {and} \bibinfo{person}{Huan Wang}.} \bibinfo{year}{2024}\natexlab{}.
\newblock \showarticletitle{Language Models are Hidden Reasoners: Unlocking Latent Reasoning Capabilities via Self-Rewarding}.
\newblock \bibinfo{journal}{\emph{CoRR}}  \bibinfo{volume}{abs/2411.04282} (\bibinfo{year}{2024}).
\newblock
\href{https://doi.org/10.48550/ARXIV.2411.04282}{doi:\nolinkurl{10.48550/ARXIV.2411.04282}}
\showeprint[arXiv]{2411.04282}


\bibitem[Chen et~al\mbox{.}(2022)]%
        {rec-denoiser}
\bibfield{author}{\bibinfo{person}{Huiyuan Chen}, \bibinfo{person}{Yusan Lin}, \bibinfo{person}{Menghai Pan}, \bibinfo{person}{Lan Wang}, \bibinfo{person}{Chin{-}Chia~Michael Yeh}, \bibinfo{person}{Xiaoting Li}, \bibinfo{person}{Yan Zheng}, \bibinfo{person}{Fei Wang}, {and} \bibinfo{person}{Hao Yang}.} \bibinfo{year}{2022}\natexlab{}.
\newblock \showarticletitle{Denoising Self-Attentive Sequential Recommendation}. In \bibinfo{booktitle}{\emph{RecSys '22: Sixteenth {ACM} Conference on Recommender Systems, Seattle, WA, USA, September 18 - 23, 2022}}, \bibfield{editor}{\bibinfo{person}{Jennifer Golbeck}, \bibinfo{person}{F.~Maxwell Harper}, \bibinfo{person}{Vanessa Murdock}, \bibinfo{person}{Michael~D. Ekstrand}, \bibinfo{person}{Bracha Shapira}, \bibinfo{person}{Justin Basilico}, \bibinfo{person}{Keld~T. Lundgaard}, {and} \bibinfo{person}{Even Oldridge}} (Eds.). \bibinfo{publisher}{{ACM}}, \bibinfo{pages}{92--101}.
\newblock
\href{https://doi.org/10.1145/3523227.3546788}{doi:\nolinkurl{10.1145/3523227.3546788}}


\bibitem[Chen et~al\mbox{.}(2025a)]%
        {still-3}
\bibfield{author}{\bibinfo{person}{Zhipeng Chen}, \bibinfo{person}{Yingqian Min}, \bibinfo{person}{Beichen Zhang}, \bibinfo{person}{Jie Chen}, \bibinfo{person}{Jinhao Jiang}, \bibinfo{person}{Daixuan Cheng}, \bibinfo{person}{Wayne~Xin Zhao}, \bibinfo{person}{Zheng Liu}, \bibinfo{person}{Xu Miao}, \bibinfo{person}{Yang Lu}, \bibinfo{person}{Lei Fang}, \bibinfo{person}{Zhongyuan Wang}, {and} \bibinfo{person}{Ji{-}Rong Wen}.} \bibinfo{year}{2025}\natexlab{a}.
\newblock \showarticletitle{An Empirical Study on Eliciting and Improving R1-like Reasoning Models}.
\newblock \bibinfo{journal}{\emph{CoRR}}  \bibinfo{volume}{abs/2503.04548} (\bibinfo{year}{2025}).
\newblock
\href{https://doi.org/10.48550/ARXIV.2503.04548}{doi:\nolinkurl{10.48550/ARXIV.2503.04548}}
\showeprint[arXiv]{2503.04548}


\bibitem[Chen et~al\mbox{.}(2025b)]%
        {r1-like}
\bibfield{author}{\bibinfo{person}{Zhipeng Chen}, \bibinfo{person}{Yingqian Min}, \bibinfo{person}{Beichen Zhang}, \bibinfo{person}{Jie Chen}, \bibinfo{person}{Jinhao Jiang}, \bibinfo{person}{Daixuan Cheng}, \bibinfo{person}{Wayne~Xin Zhao}, \bibinfo{person}{Zheng Liu}, \bibinfo{person}{Xu Miao}, \bibinfo{person}{Yang Lu}, \bibinfo{person}{Lei Fang}, \bibinfo{person}{Zhongyuan Wang}, {and} \bibinfo{person}{Ji{-}Rong Wen}.} \bibinfo{year}{2025}\natexlab{b}.
\newblock \showarticletitle{An Empirical Study on Eliciting and Improving R1-like Reasoning Models}.
\newblock \bibinfo{journal}{\emph{CoRR}}  \bibinfo{volume}{abs/2503.04548} (\bibinfo{year}{2025}).
\newblock
\href{https://doi.org/10.48550/ARXIV.2503.04548}{doi:\nolinkurl{10.48550/ARXIV.2503.04548}}
\showeprint[arXiv]{2503.04548}


\bibitem[DeepSeek{-}AI et~al\mbox{.}(2025)]%
        {deepseek-r1}
\bibfield{author}{\bibinfo{person}{DeepSeek{-}AI}, \bibinfo{person}{Daya Guo}, \bibinfo{person}{Dejian Yang}, \bibinfo{person}{Haowei Zhang}, \bibinfo{person}{Junxiao Song}, \bibinfo{person}{Ruoyu Zhang}, \bibinfo{person}{Runxin Xu}, \bibinfo{person}{Qihao Zhu}, \bibinfo{person}{Shirong Ma}, \bibinfo{person}{Peiyi Wang}, \bibinfo{person}{Xiao Bi}, \bibinfo{person}{Xiaokang Zhang}, \bibinfo{person}{Xingkai Yu}, \bibinfo{person}{Yu Wu}, \bibinfo{person}{Z.~F. Wu}, \bibinfo{person}{Zhibin Gou}, \bibinfo{person}{Zhihong Shao}, \bibinfo{person}{Zhuoshu Li}, \bibinfo{person}{Ziyi Gao}, \bibinfo{person}{Aixin Liu}, \bibinfo{person}{Bing Xue}, \bibinfo{person}{Bingxuan Wang}, \bibinfo{person}{Bochao Wu}, \bibinfo{person}{Bei Feng}, \bibinfo{person}{Chengda Lu}, \bibinfo{person}{Chenggang Zhao}, \bibinfo{person}{Chengqi Deng}, \bibinfo{person}{Chenyu Zhang}, \bibinfo{person}{Chong Ruan}, \bibinfo{person}{Damai Dai}, \bibinfo{person}{Deli Chen}, \bibinfo{person}{Dongjie Ji}, \bibinfo{person}{Erhang Li},
  \bibinfo{person}{Fangyun Lin}, \bibinfo{person}{Fucong Dai}, \bibinfo{person}{Fuli Luo}, \bibinfo{person}{Guangbo Hao}, \bibinfo{person}{Guanting Chen}, \bibinfo{person}{Guowei Li}, \bibinfo{person}{H. Zhang}, \bibinfo{person}{Han Bao}, \bibinfo{person}{Hanwei Xu}, \bibinfo{person}{Haocheng Wang}, \bibinfo{person}{Honghui Ding}, \bibinfo{person}{Huajian Xin}, \bibinfo{person}{Huazuo Gao}, \bibinfo{person}{Hui Qu}, \bibinfo{person}{Hui Li}, \bibinfo{person}{Jianzhong Guo}, \bibinfo{person}{Jiashi Li}, \bibinfo{person}{Jiawei Wang}, \bibinfo{person}{Jingchang Chen}, \bibinfo{person}{Jingyang Yuan}, \bibinfo{person}{Junjie Qiu}, \bibinfo{person}{Junlong Li}, \bibinfo{person}{J.~L. Cai}, \bibinfo{person}{Jiaqi Ni}, \bibinfo{person}{Jian Liang}, \bibinfo{person}{Jin Chen}, \bibinfo{person}{Kai Dong}, \bibinfo{person}{Kai Hu}, \bibinfo{person}{Kaige Gao}, \bibinfo{person}{Kang Guan}, \bibinfo{person}{Kexin Huang}, \bibinfo{person}{Kuai Yu}, \bibinfo{person}{Lean Wang}, \bibinfo{person}{Lecong Zhang},
  \bibinfo{person}{Liang Zhao}, \bibinfo{person}{Litong Wang}, \bibinfo{person}{Liyue Zhang}, \bibinfo{person}{Lei Xu}, \bibinfo{person}{Leyi Xia}, \bibinfo{person}{Mingchuan Zhang}, \bibinfo{person}{Minghua Zhang}, \bibinfo{person}{Minghui Tang}, \bibinfo{person}{Meng Li}, \bibinfo{person}{Miaojun Wang}, \bibinfo{person}{Mingming Li}, \bibinfo{person}{Ning Tian}, \bibinfo{person}{Panpan Huang}, \bibinfo{person}{Peng Zhang}, \bibinfo{person}{Qiancheng Wang}, \bibinfo{person}{Qinyu Chen}, \bibinfo{person}{Qiushi Du}, \bibinfo{person}{Ruiqi Ge}, \bibinfo{person}{Ruisong Zhang}, \bibinfo{person}{Ruizhe Pan}, \bibinfo{person}{Runji Wang}, \bibinfo{person}{R.~J. Chen}, \bibinfo{person}{R.~L. Jin}, \bibinfo{person}{Ruyi Chen}, \bibinfo{person}{Shanghao Lu}, \bibinfo{person}{Shangyan Zhou}, \bibinfo{person}{Shanhuang Chen}, \bibinfo{person}{Shengfeng Ye}, \bibinfo{person}{Shiyu Wang}, \bibinfo{person}{Shuiping Yu}, \bibinfo{person}{Shunfeng Zhou}, \bibinfo{person}{Shuting Pan}, {and} \bibinfo{person}{S.~S. Li}.}
  \bibinfo{year}{2025}\natexlab{}.
\newblock \showarticletitle{DeepSeek-R1: Incentivizing Reasoning Capability in LLMs via Reinforcement Learning}.
\newblock \bibinfo{journal}{\emph{CoRR}}  \bibinfo{volume}{abs/2501.12948} (\bibinfo{year}{2025}).
\newblock
\href{https://doi.org/10.48550/ARXIV.2501.12948}{doi:\nolinkurl{10.48550/ARXIV.2501.12948}}
\showeprint[arXiv]{2501.12948}


\bibitem[Geiping et~al\mbox{.}(2025)]%
        {recurrent-reasoning}
\bibfield{author}{\bibinfo{person}{Jonas Geiping}, \bibinfo{person}{Sean McLeish}, \bibinfo{person}{Neel Jain}, \bibinfo{person}{John Kirchenbauer}, \bibinfo{person}{Siddharth Singh}, \bibinfo{person}{Brian~R. Bartoldson}, \bibinfo{person}{Bhavya Kailkhura}, \bibinfo{person}{Abhinav Bhatele}, {and} \bibinfo{person}{Tom Goldstein}.} \bibinfo{year}{2025}\natexlab{}.
\newblock \showarticletitle{Scaling up Test-Time Compute with Latent Reasoning: {A} Recurrent Depth Approach}.
\newblock \bibinfo{journal}{\emph{CoRR}}  \bibinfo{volume}{abs/2502.05171} (\bibinfo{year}{2025}).
\newblock
\href{https://doi.org/10.48550/ARXIV.2502.05171}{doi:\nolinkurl{10.48550/ARXIV.2502.05171}}
\showeprint[arXiv]{2502.05171}


\bibitem[Guo et~al\mbox{.}(2024)]%
        {deepseek-coder}
\bibfield{author}{\bibinfo{person}{Daya Guo}, \bibinfo{person}{Qihao Zhu}, \bibinfo{person}{Dejian Yang}, \bibinfo{person}{Zhenda Xie}, \bibinfo{person}{Kai Dong}, \bibinfo{person}{Wentao Zhang}, \bibinfo{person}{Guanting Chen}, \bibinfo{person}{Xiao Bi}, \bibinfo{person}{Y. Wu}, \bibinfo{person}{Y.~K. Li}, \bibinfo{person}{Fuli Luo}, \bibinfo{person}{Yingfei Xiong}, {and} \bibinfo{person}{Wenfeng Liang}.} \bibinfo{year}{2024}\natexlab{}.
\newblock \showarticletitle{DeepSeek-Coder: When the Large Language Model Meets Programming - The Rise of Code Intelligence}.
\newblock \bibinfo{journal}{\emph{CoRR}}  \bibinfo{volume}{abs/2401.14196} (\bibinfo{year}{2024}).
\newblock
\href{https://doi.org/10.48550/ARXIV.2401.14196}{doi:\nolinkurl{10.48550/ARXIV.2401.14196}}
\showeprint[arXiv]{2401.14196}


\bibitem[Hao et~al\mbox{.}(2024)]%
        {coconut}
\bibfield{author}{\bibinfo{person}{Shibo Hao}, \bibinfo{person}{Sainbayar Sukhbaatar}, \bibinfo{person}{DiJia Su}, \bibinfo{person}{Xian Li}, \bibinfo{person}{Zhiting Hu}, \bibinfo{person}{Jason Weston}, {and} \bibinfo{person}{Yuandong Tian}.} \bibinfo{year}{2024}\natexlab{}.
\newblock \showarticletitle{Training Large Language Models to Reason in a Continuous Latent Space}.
\newblock \bibinfo{journal}{\emph{CoRR}}  \bibinfo{volume}{abs/2412.06769} (\bibinfo{year}{2024}).
\newblock
\href{https://doi.org/10.48550/ARXIV.2412.06769}{doi:\nolinkurl{10.48550/ARXIV.2412.06769}}
\showeprint[arXiv]{2412.06769}


\bibitem[Hao et~al\mbox{.}(2023)]%
        {DBLP:journals/tkde/HaoZZLSXLZ23}
\bibfield{author}{\bibinfo{person}{Yongjing Hao}, \bibinfo{person}{Tingting Zhang}, \bibinfo{person}{Pengpeng Zhao}, \bibinfo{person}{Yanchi Liu}, \bibinfo{person}{Victor~S. Sheng}, \bibinfo{person}{Jiajie Xu}, \bibinfo{person}{Guanfeng Liu}, {and} \bibinfo{person}{Xiaofang Zhou}.} \bibinfo{year}{2023}\natexlab{}.
\newblock \showarticletitle{Feature-Level Deeper Self-Attention Network With Contrastive Learning for Sequential Recommendation}.
\newblock \bibinfo{journal}{\emph{{IEEE} Trans. Knowl. Data Eng.}} \bibinfo{volume}{35}, \bibinfo{number}{10} (\bibinfo{year}{2023}), \bibinfo{pages}{10112--10124}.
\newblock
\href{https://doi.org/10.1109/TKDE.2023.3250463}{doi:\nolinkurl{10.1109/TKDE.2023.3250463}}


\bibitem[He and McAuley(2016)]%
        {fossil}
\bibfield{author}{\bibinfo{person}{Ruining He} {and} \bibinfo{person}{Julian~J. McAuley}.} \bibinfo{year}{2016}\natexlab{}.
\newblock \showarticletitle{Fusing Similarity Models with Markov Chains for Sparse Sequential Recommendation}. In \bibinfo{booktitle}{\emph{{IEEE} 16th International Conference on Data Mining, {ICDM} 2016, December 12-15, 2016, Barcelona, Spain}}, \bibfield{editor}{\bibinfo{person}{Francesco Bonchi}, \bibinfo{person}{Josep Domingo{-}Ferrer}, \bibinfo{person}{Ricardo Baeza{-}Yates}, \bibinfo{person}{Zhi{-}Hua Zhou}, {and} \bibinfo{person}{Xindong Wu}} (Eds.). \bibinfo{publisher}{{IEEE} Computer Society}, \bibinfo{pages}{191--200}.
\newblock
\href{https://doi.org/10.1109/ICDM.2016.0030}{doi:\nolinkurl{10.1109/ICDM.2016.0030}}


\bibitem[Hidasi et~al\mbox{.}(2016a)]%
        {gru4rec}
\bibfield{author}{\bibinfo{person}{Bal{\'{a}}zs Hidasi}, \bibinfo{person}{Alexandros Karatzoglou}, \bibinfo{person}{Linas Baltrunas}, {and} \bibinfo{person}{Domonkos Tikk}.} \bibinfo{year}{2016}\natexlab{a}.
\newblock \showarticletitle{Session-based Recommendations with Recurrent Neural Networks}. In \bibinfo{booktitle}{\emph{4th International Conference on Learning Representations, {ICLR} 2016, San Juan, Puerto Rico, May 2-4, 2016, Conference Track Proceedings}}, \bibfield{editor}{\bibinfo{person}{Yoshua Bengio} {and} \bibinfo{person}{Yann LeCun}} (Eds.).
\newblock
\urldef\tempurl%
\url{http://arxiv.org/abs/1511.06939}
\showURL{%
\tempurl}


\bibitem[Hidasi et~al\mbox{.}(2016b)]%
        {DBLP:conf/recsys/HidasiQKT16}
\bibfield{author}{\bibinfo{person}{Bal{\'{a}}zs Hidasi}, \bibinfo{person}{Massimo Quadrana}, \bibinfo{person}{Alexandros Karatzoglou}, {and} \bibinfo{person}{Domonkos Tikk}.} \bibinfo{year}{2016}\natexlab{b}.
\newblock \showarticletitle{Parallel Recurrent Neural Network Architectures for Feature-rich Session-based Recommendations}. In \bibinfo{booktitle}{\emph{Proceedings of the 10th {ACM} Conference on Recommender Systems, Boston, MA, USA, September 15-19, 2016}}, \bibfield{editor}{\bibinfo{person}{Shilad Sen}, \bibinfo{person}{Werner Geyer}, \bibinfo{person}{Jill Freyne}, {and} \bibinfo{person}{Pablo Castells}} (Eds.). \bibinfo{publisher}{{ACM}}, \bibinfo{pages}{241--248}.
\newblock
\href{https://doi.org/10.1145/2959100.2959167}{doi:\nolinkurl{10.1145/2959100.2959167}}


\bibitem[Hou et~al\mbox{.}(2022)]%
        {unisrec}
\bibfield{author}{\bibinfo{person}{Yupeng Hou}, \bibinfo{person}{Shanlei Mu}, \bibinfo{person}{Wayne~Xin Zhao}, \bibinfo{person}{Yaliang Li}, \bibinfo{person}{Bolin Ding}, {and} \bibinfo{person}{Ji{-}Rong Wen}.} \bibinfo{year}{2022}\natexlab{}.
\newblock \showarticletitle{Towards Universal Sequence Representation Learning for Recommender Systems}. In \bibinfo{booktitle}{\emph{{KDD} '22: The 28th {ACM} {SIGKDD} Conference on Knowledge Discovery and Data Mining, Washington, DC, USA, August 14 - 18, 2022}}. \bibinfo{publisher}{{ACM}}, \bibinfo{pages}{585--593}.
\newblock
\href{https://doi.org/10.1145/3534678.3539381}{doi:\nolinkurl{10.1145/3534678.3539381}}


\bibitem[Jaech et~al\mbox{.}(2024)]%
        {openai-o1}
\bibfield{author}{\bibinfo{person}{Aaron Jaech}, \bibinfo{person}{Adam Kalai}, \bibinfo{person}{Adam Lerer}, \bibinfo{person}{Adam Richardson}, \bibinfo{person}{Ahmed El{-}Kishky}, \bibinfo{person}{Aiden Low}, \bibinfo{person}{Alec Helyar}, \bibinfo{person}{Aleksander Madry}, \bibinfo{person}{Alex Beutel}, \bibinfo{person}{Alex Carney}, \bibinfo{person}{Alex Iftimie}, \bibinfo{person}{Alex Karpenko}, \bibinfo{person}{Alex~Tachard Passos}, \bibinfo{person}{Alexander Neitz}, \bibinfo{person}{Alexander Prokofiev}, \bibinfo{person}{Alexander Wei}, \bibinfo{person}{Allison Tam}, \bibinfo{person}{Ally Bennett}, \bibinfo{person}{Ananya Kumar}, \bibinfo{person}{Andre Saraiva}, \bibinfo{person}{Andrea Vallone}, \bibinfo{person}{Andrew Duberstein}, \bibinfo{person}{Andrew Kondrich}, \bibinfo{person}{Andrey Mishchenko}, \bibinfo{person}{Andy Applebaum}, \bibinfo{person}{Angela Jiang}, \bibinfo{person}{Ashvin Nair}, \bibinfo{person}{Barret Zoph}, \bibinfo{person}{Behrooz Ghorbani}, \bibinfo{person}{Ben Rossen},
  \bibinfo{person}{Benjamin Sokolowsky}, \bibinfo{person}{Boaz Barak}, \bibinfo{person}{Bob McGrew}, \bibinfo{person}{Borys Minaiev}, \bibinfo{person}{Botao Hao}, \bibinfo{person}{Bowen Baker}, \bibinfo{person}{Brandon Houghton}, \bibinfo{person}{Brandon McKinzie}, \bibinfo{person}{Brydon Eastman}, \bibinfo{person}{Camillo Lugaresi}, \bibinfo{person}{Cary Bassin}, \bibinfo{person}{Cary Hudson}, \bibinfo{person}{Chak~Ming Li}, \bibinfo{person}{Charles de Bourcy}, \bibinfo{person}{Chelsea Voss}, \bibinfo{person}{Chen Shen}, \bibinfo{person}{Chong Zhang}, \bibinfo{person}{Chris Koch}, \bibinfo{person}{Chris Orsinger}, \bibinfo{person}{Christopher Hesse}, \bibinfo{person}{Claudia Fischer}, \bibinfo{person}{Clive Chan}, \bibinfo{person}{Dan Roberts}, \bibinfo{person}{Daniel Kappler}, \bibinfo{person}{Daniel Levy}, \bibinfo{person}{Daniel Selsam}, \bibinfo{person}{David Dohan}, \bibinfo{person}{David Farhi}, \bibinfo{person}{David Mely}, \bibinfo{person}{David Robinson}, \bibinfo{person}{Dimitris Tsipras},
  \bibinfo{person}{Doug Li}, \bibinfo{person}{Dragos Oprica}, \bibinfo{person}{Eben Freeman}, \bibinfo{person}{Eddie Zhang}, \bibinfo{person}{Edmund Wong}, \bibinfo{person}{Elizabeth Proehl}, \bibinfo{person}{Enoch Cheung}, \bibinfo{person}{Eric Mitchell}, \bibinfo{person}{Eric Wallace}, \bibinfo{person}{Erik Ritter}, \bibinfo{person}{Evan Mays}, \bibinfo{person}{Fan Wang}, \bibinfo{person}{Felipe~Petroski Such}, \bibinfo{person}{Filippo Raso}, \bibinfo{person}{Florencia Leoni}, \bibinfo{person}{Foivos Tsimpourlas}, \bibinfo{person}{Francis Song}, \bibinfo{person}{Fred von Lohmann}, \bibinfo{person}{Freddie Sulit}, \bibinfo{person}{Geoff Salmon}, \bibinfo{person}{Giambattista Parascandolo}, \bibinfo{person}{Gildas Chabot}, \bibinfo{person}{Grace Zhao}, \bibinfo{person}{Greg Brockman}, \bibinfo{person}{Guillaume Leclerc}, \bibinfo{person}{Hadi Salman}, \bibinfo{person}{Haiming Bao}, \bibinfo{person}{Hao Sheng}, \bibinfo{person}{Hart Andrin}, \bibinfo{person}{Hessam Bagherinezhad}, \bibinfo{person}{Hongyu Ren},
  \bibinfo{person}{Hunter Lightman}, \bibinfo{person}{Hyung~Won Chung}, \bibinfo{person}{Ian Kivlichan}, \bibinfo{person}{Ian O'Connell}, \bibinfo{person}{Ian Osband}, \bibinfo{person}{Ignasi~Clavera Gilaberte}, {and} \bibinfo{person}{Ilge Akkaya}.} \bibinfo{year}{2024}\natexlab{}.
\newblock \showarticletitle{OpenAI o1 System Card}.
\newblock \bibinfo{journal}{\emph{CoRR}}  \bibinfo{volume}{abs/2412.16720} (\bibinfo{year}{2024}).
\newblock
\href{https://doi.org/10.48550/ARXIV.2412.16720}{doi:\nolinkurl{10.48550/ARXIV.2412.16720}}
\showeprint[arXiv]{2412.16720}


\bibitem[Kang and McAuley(2018)]%
        {sasrec}
\bibfield{author}{\bibinfo{person}{Wang{-}Cheng Kang} {and} \bibinfo{person}{Julian~J. McAuley}.} \bibinfo{year}{2018}\natexlab{}.
\newblock \showarticletitle{Self-Attentive Sequential Recommendation}. In \bibinfo{booktitle}{\emph{{IEEE} International Conference on Data Mining, {ICDM} 2018, Singapore, November 17-20, 2018}}. \bibinfo{publisher}{{IEEE} Computer Society}, \bibinfo{pages}{197--206}.
\newblock
\href{https://doi.org/10.1109/ICDM.2018.00035}{doi:\nolinkurl{10.1109/ICDM.2018.00035}}


\bibitem[Lightman et~al\mbox{.}(2024)]%
        {verify-step-by-step}
\bibfield{author}{\bibinfo{person}{Hunter Lightman}, \bibinfo{person}{Vineet Kosaraju}, \bibinfo{person}{Yuri Burda}, \bibinfo{person}{Harrison Edwards}, \bibinfo{person}{Bowen Baker}, \bibinfo{person}{Teddy Lee}, \bibinfo{person}{Jan Leike}, \bibinfo{person}{John Schulman}, \bibinfo{person}{Ilya Sutskever}, {and} \bibinfo{person}{Karl Cobbe}.} \bibinfo{year}{2024}\natexlab{}.
\newblock \showarticletitle{Let's Verify Step by Step}. In \bibinfo{booktitle}{\emph{The Twelfth International Conference on Learning Representations, {ICLR} 2024, Vienna, Austria, May 7-11, 2024}}. \bibinfo{publisher}{OpenReview.net}.
\newblock
\urldef\tempurl%
\url{https://openreview.net/forum?id=v8L0pN6EOi}
\showURL{%
\tempurl}


\bibitem[Liu et~al\mbox{.}(2025)]%
        {cocorec}
\bibfield{author}{\bibinfo{person}{Enze Liu}, \bibinfo{person}{Bowen Zheng}, \bibinfo{person}{Wayne~Xin Zhao}, {and} \bibinfo{person}{Ji{-}Rong Wen}.} \bibinfo{year}{2025}\natexlab{}.
\newblock \showarticletitle{Bridging Textual-Collaborative Gap through Semantic Codes for Sequential Recommendation}.
\newblock \bibinfo{journal}{\emph{CoRR}}  \bibinfo{volume}{abs/2503.12183} (\bibinfo{year}{2025}).
\newblock
\href{https://doi.org/10.48550/ARXIV.2503.12183}{doi:\nolinkurl{10.48550/ARXIV.2503.12183}}
\showeprint[arXiv]{2503.12183}


\bibitem[Liu et~al\mbox{.}(2024)]%
        {fame}
\bibfield{author}{\bibinfo{person}{Mingrui Liu}, \bibinfo{person}{Sixiao Zhang}, {and} \bibinfo{person}{Cheng Long}.} \bibinfo{year}{2024}\natexlab{}.
\newblock \showarticletitle{Facet-Aware Multi-Head Mixture-of-Experts Model for Sequential Recommendation}.
\newblock \bibinfo{journal}{\emph{CoRR}}  \bibinfo{volume}{abs/2411.01457} (\bibinfo{year}{2024}).
\newblock
\href{https://doi.org/10.48550/ARXIV.2411.01457}{doi:\nolinkurl{10.48550/ARXIV.2411.01457}}
\showeprint[arXiv]{2411.01457}


\bibitem[Luo et~al\mbox{.}(2024)]%
        {omegaprm}
\bibfield{author}{\bibinfo{person}{Liangchen Luo}, \bibinfo{person}{Yinxiao Liu}, \bibinfo{person}{Rosanne Liu}, \bibinfo{person}{Samrat Phatale}, \bibinfo{person}{Harsh Lara}, \bibinfo{person}{Yunxuan Li}, \bibinfo{person}{Lei Shu}, \bibinfo{person}{Yun Zhu}, \bibinfo{person}{Lei Meng}, \bibinfo{person}{Jiao Sun}, {and} \bibinfo{person}{Abhinav Rastogi}.} \bibinfo{year}{2024}\natexlab{}.
\newblock \showarticletitle{Improve Mathematical Reasoning in Language Models by Automated Process Supervision}.
\newblock \bibinfo{journal}{\emph{CoRR}}  \bibinfo{volume}{abs/2406.06592} (\bibinfo{year}{2024}).
\newblock
\href{https://doi.org/10.48550/ARXIV.2406.06592}{doi:\nolinkurl{10.48550/ARXIV.2406.06592}}
\showeprint[arXiv]{2406.06592}


\bibitem[Qiu et~al\mbox{.}(2022)]%
        {duorec}
\bibfield{author}{\bibinfo{person}{Ruihong Qiu}, \bibinfo{person}{Zi Huang}, \bibinfo{person}{Hongzhi Yin}, {and} \bibinfo{person}{Zijian Wang}.} \bibinfo{year}{2022}\natexlab{}.
\newblock \showarticletitle{Contrastive Learning for Representation Degeneration Problem in Sequential Recommendation}. In \bibinfo{booktitle}{\emph{{WSDM} '22: The Fifteenth {ACM} International Conference on Web Search and Data Mining, Virtual Event / Tempe, AZ, USA, February 21 - 25, 2022}}, \bibfield{editor}{\bibinfo{person}{K.~Selcuk Candan}, \bibinfo{person}{Huan Liu}, \bibinfo{person}{Leman Akoglu}, \bibinfo{person}{Xin~Luna Dong}, {and} \bibinfo{person}{Jiliang Tang}} (Eds.). \bibinfo{publisher}{{ACM}}, \bibinfo{pages}{813--823}.
\newblock
\href{https://doi.org/10.1145/3488560.3498433}{doi:\nolinkurl{10.1145/3488560.3498433}}


\bibitem[Quadrana et~al\mbox{.}(2017)]%
        {hrnn}
\bibfield{author}{\bibinfo{person}{Massimo Quadrana}, \bibinfo{person}{Alexandros Karatzoglou}, \bibinfo{person}{Bal{\'{a}}zs Hidasi}, {and} \bibinfo{person}{Paolo Cremonesi}.} \bibinfo{year}{2017}\natexlab{}.
\newblock \showarticletitle{Personalizing Session-based Recommendations with Hierarchical Recurrent Neural Networks}. In \bibinfo{booktitle}{\emph{Proceedings of the Eleventh {ACM} Conference on Recommender Systems, RecSys 2017, Como, Italy, August 27-31, 2017}}, \bibfield{editor}{\bibinfo{person}{Paolo Cremonesi}, \bibinfo{person}{Francesco Ricci}, \bibinfo{person}{Shlomo Berkovsky}, {and} \bibinfo{person}{Alexander Tuzhilin}} (Eds.). \bibinfo{publisher}{{ACM}}, \bibinfo{pages}{130--137}.
\newblock
\href{https://doi.org/10.1145/3109859.3109896}{doi:\nolinkurl{10.1145/3109859.3109896}}


\bibitem[Rajput et~al\mbox{.}(2023)]%
        {tiger}
\bibfield{author}{\bibinfo{person}{Shashank Rajput}, \bibinfo{person}{Nikhil Mehta}, \bibinfo{person}{Anima Singh}, \bibinfo{person}{Raghunandan~Hulikal Keshavan}, \bibinfo{person}{Trung Vu}, \bibinfo{person}{Lukasz Heldt}, \bibinfo{person}{Lichan Hong}, \bibinfo{person}{Yi Tay}, \bibinfo{person}{Vinh~Q. Tran}, \bibinfo{person}{Jonah Samost}, \bibinfo{person}{Maciej Kula}, \bibinfo{person}{Ed~H. Chi}, {and} \bibinfo{person}{Mahesh Sathiamoorthy}.} \bibinfo{year}{2023}\natexlab{}.
\newblock \showarticletitle{Recommender Systems with Generative Retrieval}. In \bibinfo{booktitle}{\emph{Advances in Neural Information Processing Systems 36: Annual Conference on Neural Information Processing Systems 2023, NeurIPS 2023, New Orleans, LA, USA, December 10 - 16, 2023}}.
\newblock
\urldef\tempurl%
\url{http://papers.nips.cc/paper\_files/paper/2023/hash/20dcab0f14046a5c6b02b61da9f13229-Abstract-Conference.html}
\showURL{%
\tempurl}


\bibitem[Rendle et~al\mbox{.}(2010)]%
        {fpmc}
\bibfield{author}{\bibinfo{person}{Steffen Rendle}, \bibinfo{person}{Christoph Freudenthaler}, {and} \bibinfo{person}{Lars Schmidt{-}Thieme}.} \bibinfo{year}{2010}\natexlab{}.
\newblock \showarticletitle{Factorizing personalized Markov chains for next-basket recommendation}. In \bibinfo{booktitle}{\emph{Proceedings of the 19th International Conference on World Wide Web, {WWW} 2010, Raleigh, North Carolina, USA, April 26-30, 2010}}, \bibfield{editor}{\bibinfo{person}{Michael Rappa}, \bibinfo{person}{Paul Jones}, \bibinfo{person}{Juliana Freire}, {and} \bibinfo{person}{Soumen Chakrabarti}} (Eds.). \bibinfo{publisher}{{ACM}}, \bibinfo{pages}{811--820}.
\newblock
\href{https://doi.org/10.1145/1772690.1772773}{doi:\nolinkurl{10.1145/1772690.1772773}}


\bibitem[Saunshi et~al\mbox{.}(2025)]%
        {loop-power}
\bibfield{author}{\bibinfo{person}{Nikunj Saunshi}, \bibinfo{person}{Nishanth Dikkala}, \bibinfo{person}{Zhiyuan Li}, \bibinfo{person}{Sanjiv Kumar}, {and} \bibinfo{person}{Sashank~J. Reddi}.} \bibinfo{year}{2025}\natexlab{}.
\newblock \showarticletitle{Reasoning with Latent Thoughts: On the Power of Looped Transformers}.
\newblock \bibinfo{journal}{\emph{CoRR}}  \bibinfo{volume}{abs/2502.17416} (\bibinfo{year}{2025}).
\newblock
\href{https://doi.org/10.48550/ARXIV.2502.17416}{doi:\nolinkurl{10.48550/ARXIV.2502.17416}}
\showeprint[arXiv]{2502.17416}


\bibitem[Shao et~al\mbox{.}(2024)]%
        {grpo}
\bibfield{author}{\bibinfo{person}{Zhihong Shao}, \bibinfo{person}{Peiyi Wang}, \bibinfo{person}{Qihao Zhu}, \bibinfo{person}{Runxin Xu}, \bibinfo{person}{Junxiao Song}, \bibinfo{person}{Mingchuan Zhang}, \bibinfo{person}{Y.~K. Li}, \bibinfo{person}{Y. Wu}, {and} \bibinfo{person}{Daya Guo}.} \bibinfo{year}{2024}\natexlab{}.
\newblock \showarticletitle{DeepSeekMath: Pushing the Limits of Mathematical Reasoning in Open Language Models}.
\newblock \bibinfo{journal}{\emph{CoRR}}  \bibinfo{volume}{abs/2402.03300} (\bibinfo{year}{2024}).
\newblock
\href{https://doi.org/10.48550/ARXIV.2402.03300}{doi:\nolinkurl{10.48550/ARXIV.2402.03300}}
\showeprint[arXiv]{2402.03300}


\bibitem[Shin et~al\mbox{.}(2024)]%
        {bsarec}
\bibfield{author}{\bibinfo{person}{Yehjin Shin}, \bibinfo{person}{Jeongwhan Choi}, \bibinfo{person}{Hyowon Wi}, {and} \bibinfo{person}{Noseong Park}.} \bibinfo{year}{2024}\natexlab{}.
\newblock \showarticletitle{An Attentive Inductive Bias for Sequential Recommendation beyond the Self-Attention}. In \bibinfo{booktitle}{\emph{Thirty-Eighth {AAAI} Conference on Artificial Intelligence, {AAAI} 2024, Thirty-Sixth Conference on Innovative Applications of Artificial Intelligence, {IAAI} 2024, Fourteenth Symposium on Educational Advances in Artificial Intelligence, {EAAI} 2014, February 20-27, 2024, Vancouver, Canada}}, \bibfield{editor}{\bibinfo{person}{Michael~J. Wooldridge}, \bibinfo{person}{Jennifer~G. Dy}, {and} \bibinfo{person}{Sriraam Natarajan}} (Eds.). \bibinfo{publisher}{{AAAI} Press}, \bibinfo{pages}{8984--8992}.
\newblock
\href{https://doi.org/10.1609/AAAI.V38I8.28747}{doi:\nolinkurl{10.1609/AAAI.V38I8.28747}}


\bibitem[Singer et~al\mbox{.}(2022)]%
        {e-commerce}
\bibfield{author}{\bibinfo{person}{Uriel Singer}, \bibinfo{person}{Haggai Roitman}, \bibinfo{person}{Yotam Eshel}, \bibinfo{person}{Alexander Nus}, \bibinfo{person}{Ido Guy}, \bibinfo{person}{Or Levi}, \bibinfo{person}{Idan Hasson}, {and} \bibinfo{person}{Eliyahu Kiperwasser}.} \bibinfo{year}{2022}\natexlab{}.
\newblock \showarticletitle{Sequential Modeling with Multiple Attributes for Watchlist Recommendation in E-Commerce}. In \bibinfo{booktitle}{\emph{{WSDM} '22: The Fifteenth {ACM} International Conference on Web Search and Data Mining, Virtual Event / Tempe, AZ, USA, February 21 - 25, 2022}}, \bibfield{editor}{\bibinfo{person}{K.~Selcuk Candan}, \bibinfo{person}{Huan Liu}, \bibinfo{person}{Leman Akoglu}, \bibinfo{person}{Xin~Luna Dong}, {and} \bibinfo{person}{Jiliang Tang}} (Eds.). \bibinfo{publisher}{{ACM}}, \bibinfo{pages}{937--946}.
\newblock
\href{https://doi.org/10.1145/3488560.3498453}{doi:\nolinkurl{10.1145/3488560.3498453}}


\bibitem[Su et~al\mbox{.}(2025)]%
        {overthinking}
\bibfield{author}{\bibinfo{person}{Jinyan Su}, \bibinfo{person}{Jennifer Healey}, \bibinfo{person}{Preslav Nakov}, {and} \bibinfo{person}{Claire Cardie}.} \bibinfo{year}{2025}\natexlab{}.
\newblock \bibinfo{title}{Between Underthinking and Overthinking: An Empirical Study of Reasoning Length and correctness in LLMs}.
\newblock
\showeprint[arxiv]{2505.00127}~[cs.CL]
\urldef\tempurl%
\url{https://arxiv.org/abs/2505.00127}
\showURL{%
\tempurl}


\bibitem[Sun et~al\mbox{.}(2019)]%
        {bert4rec}
\bibfield{author}{\bibinfo{person}{Fei Sun}, \bibinfo{person}{Jun Liu}, \bibinfo{person}{Jian Wu}, \bibinfo{person}{Changhua Pei}, \bibinfo{person}{Xiao Lin}, \bibinfo{person}{Wenwu Ou}, {and} \bibinfo{person}{Peng Jiang}.} \bibinfo{year}{2019}\natexlab{}.
\newblock \showarticletitle{BERT4Rec: Sequential Recommendation with Bidirectional Encoder Representations from Transformer}. In \bibinfo{booktitle}{\emph{Proceedings of the 28th {ACM} International Conference on Information and Knowledge Management, {CIKM} 2019, Beijing, China, November 3-7, 2019}}, \bibfield{editor}{\bibinfo{person}{Wenwu Zhu}, \bibinfo{person}{Dacheng Tao}, \bibinfo{person}{Xueqi Cheng}, \bibinfo{person}{Peng Cui}, \bibinfo{person}{Elke~A. Rundensteiner}, \bibinfo{person}{David Carmel}, \bibinfo{person}{Qi~He}, {and} \bibinfo{person}{Jeffrey~Xu Yu}} (Eds.). \bibinfo{publisher}{{ACM}}, \bibinfo{pages}{1441--1450}.
\newblock
\href{https://doi.org/10.1145/3357384.3357895}{doi:\nolinkurl{10.1145/3357384.3357895}}


\bibitem[Tan et~al\mbox{.}(2016)]%
        {DBLP:conf/recsys/TanXL16}
\bibfield{author}{\bibinfo{person}{Yong~Kiam Tan}, \bibinfo{person}{Xinxing Xu}, {and} \bibinfo{person}{Yong Liu}.} \bibinfo{year}{2016}\natexlab{}.
\newblock \showarticletitle{Improved Recurrent Neural Networks for Session-based Recommendations}. In \bibinfo{booktitle}{\emph{Proceedings of the 1st Workshop on Deep Learning for Recommender Systems, DLRS@RecSys 2016, Boston, MA, USA, September 15, 2016}}. \bibinfo{publisher}{{ACM}}, \bibinfo{pages}{17--22}.
\newblock
\href{https://doi.org/10.1145/2988450.2988452}{doi:\nolinkurl{10.1145/2988450.2988452}}


\bibitem[Tang et~al\mbox{.}(2025)]%
        {rearec}
\bibfield{author}{\bibinfo{person}{Jiakai Tang}, \bibinfo{person}{Sunhao Dai}, \bibinfo{person}{Teng Shi}, \bibinfo{person}{Jun Xu}, \bibinfo{person}{Xu Chen}, \bibinfo{person}{Wen Chen}, \bibinfo{person}{Wu Jian}, {and} \bibinfo{person}{Yuning Jiang}.} \bibinfo{year}{2025}\natexlab{}.
\newblock \showarticletitle{Think Before Recommend: Unleashing the Latent Reasoning Power for Sequential Recommendation}.
\newblock \bibinfo{journal}{\emph{CoRR}}  \bibinfo{volume}{abs/2503.22675} (\bibinfo{year}{2025}).
\newblock
\href{https://doi.org/10.48550/ARXIV.2503.22675}{doi:\nolinkurl{10.48550/ARXIV.2503.22675}}
\showeprint[arXiv]{2503.22675}


\bibitem[Tang and Wang(2018)]%
        {caser}
\bibfield{author}{\bibinfo{person}{Jiaxi Tang} {and} \bibinfo{person}{Ke Wang}.} \bibinfo{year}{2018}\natexlab{}.
\newblock \showarticletitle{Personalized Top-N Sequential Recommendation via Convolutional Sequence Embedding}. In \bibinfo{booktitle}{\emph{Proceedings of the Eleventh {ACM} International Conference on Web Search and Data Mining, {WSDM} 2018, Marina Del Rey, CA, USA, February 5-9, 2018}}, \bibfield{editor}{\bibinfo{person}{Yi~Chang}, \bibinfo{person}{Chengxiang Zhai}, \bibinfo{person}{Yan Liu}, {and} \bibinfo{person}{Yoelle Maarek}} (Eds.). \bibinfo{publisher}{{ACM}}, \bibinfo{pages}{565--573}.
\newblock
\href{https://doi.org/10.1145/3159652.3159656}{doi:\nolinkurl{10.1145/3159652.3159656}}


\bibitem[Team et~al\mbox{.}(2025)]%
        {kimi-k1.5}
\bibfield{author}{\bibinfo{person}{Kimi Team}, \bibinfo{person}{Angang Du}, \bibinfo{person}{Bofei Gao}, \bibinfo{person}{Bowei Xing}, \bibinfo{person}{Changjiu Jiang}, \bibinfo{person}{Cheng Chen}, \bibinfo{person}{Cheng Li}, \bibinfo{person}{Chenjun Xiao}, \bibinfo{person}{Chenzhuang Du}, \bibinfo{person}{Chonghua Liao}, \bibinfo{person}{Chuning Tang}, \bibinfo{person}{Congcong Wang}, \bibinfo{person}{Dehao Zhang}, \bibinfo{person}{Enming Yuan}, \bibinfo{person}{Enzhe Lu}, \bibinfo{person}{Fengxiang Tang}, \bibinfo{person}{Flood Sung}, \bibinfo{person}{Guangda Wei}, \bibinfo{person}{Guokun Lai}, \bibinfo{person}{Haiqing Guo}, \bibinfo{person}{Han Zhu}, \bibinfo{person}{Hao Ding}, \bibinfo{person}{Hao Hu}, \bibinfo{person}{Hao Yang}, \bibinfo{person}{Hao Zhang}, \bibinfo{person}{Haotian Yao}, \bibinfo{person}{Haotian Zhao}, \bibinfo{person}{Haoyu Lu}, \bibinfo{person}{Haoze Li}, \bibinfo{person}{Haozhen Yu}, \bibinfo{person}{Hongcheng Gao}, \bibinfo{person}{Huabin Zheng}, \bibinfo{person}{Huan Yuan},
  \bibinfo{person}{Jia Chen}, \bibinfo{person}{Jianhang Guo}, \bibinfo{person}{Jianlin Su}, \bibinfo{person}{Jianzhou Wang}, \bibinfo{person}{Jie Zhao}, \bibinfo{person}{Jin Zhang}, \bibinfo{person}{Jingyuan Liu}, \bibinfo{person}{Junjie Yan}, \bibinfo{person}{Junyan Wu}, \bibinfo{person}{Lidong Shi}, \bibinfo{person}{Ling Ye}, \bibinfo{person}{Longhui Yu}, \bibinfo{person}{Mengnan Dong}, \bibinfo{person}{Neo Zhang}, \bibinfo{person}{Ningchen Ma}, \bibinfo{person}{Qiwei Pan}, \bibinfo{person}{Qucheng Gong}, \bibinfo{person}{Shaowei Liu}, \bibinfo{person}{Shengling Ma}, \bibinfo{person}{Shupeng Wei}, \bibinfo{person}{Sihan Cao}, \bibinfo{person}{Siying Huang}, \bibinfo{person}{Tao Jiang}, \bibinfo{person}{Weihao Gao}, \bibinfo{person}{Weimin Xiong}, \bibinfo{person}{Weiran He}, \bibinfo{person}{Weixiao Huang}, \bibinfo{person}{Wenhao Wu}, \bibinfo{person}{Wenyang He}, \bibinfo{person}{Xianghui Wei}, \bibinfo{person}{Xianqing Jia}, \bibinfo{person}{Xingzhe Wu}, \bibinfo{person}{Xinran Xu},
  \bibinfo{person}{Xinxing Zu}, \bibinfo{person}{Xinyu Zhou}, \bibinfo{person}{Xuehai Pan}, \bibinfo{person}{Y. Charles}, \bibinfo{person}{Yang Li}, \bibinfo{person}{Yangyang Hu}, \bibinfo{person}{Yangyang Liu}, \bibinfo{person}{Yanru Chen}, \bibinfo{person}{Yejie Wang}, \bibinfo{person}{Yibo Liu}, \bibinfo{person}{Yidao Qin}, \bibinfo{person}{Yifeng Liu}, \bibinfo{person}{Ying Yang}, \bibinfo{person}{Yiping Bao}, \bibinfo{person}{Yulun Du}, \bibinfo{person}{Yuxin Wu}, \bibinfo{person}{Yuzhi Wang}, \bibinfo{person}{Zaida Zhou}, \bibinfo{person}{Zhaoji Wang}, \bibinfo{person}{Zhaowei Li}, \bibinfo{person}{Zhen Zhu}, \bibinfo{person}{Zheng Zhang}, \bibinfo{person}{Zhexu Wang}, \bibinfo{person}{Zhilin Yang}, \bibinfo{person}{Zhiqi Huang}, \bibinfo{person}{Zihao Huang}, \bibinfo{person}{Ziyao Xu}, {and} \bibinfo{person}{Zonghan Yang}.} \bibinfo{year}{2025}\natexlab{}.
\newblock \showarticletitle{Kimi k1.5: Scaling Reinforcement Learning with LLMs}.
\newblock \bibinfo{journal}{\emph{CoRR}}  \bibinfo{volume}{abs/2501.12599} (\bibinfo{year}{2025}).
\newblock
\href{https://doi.org/10.48550/ARXIV.2501.12599}{doi:\nolinkurl{10.48550/ARXIV.2501.12599}}
\showeprint[arXiv]{2501.12599}


\bibitem[Wang et~al\mbox{.}(2023)]%
        {convformer}
\bibfield{author}{\bibinfo{person}{Hao Wang}, \bibinfo{person}{Jianxun Lian}, \bibinfo{person}{Mingqi Wu}, \bibinfo{person}{Haoxuan Li}, \bibinfo{person}{Jiajun Fan}, \bibinfo{person}{Wanyue Xu}, \bibinfo{person}{Chaozhuo Li}, {and} \bibinfo{person}{Xing Xie}.} \bibinfo{year}{2023}\natexlab{}.
\newblock \showarticletitle{ConvFormer: Revisiting Transformer for Sequential User Modeling}.
\newblock \bibinfo{journal}{\emph{CoRR}}  \bibinfo{volume}{abs/2308.02925} (\bibinfo{year}{2023}).
\newblock
\href{https://doi.org/10.48550/ARXIV.2308.02925}{doi:\nolinkurl{10.48550/ARXIV.2308.02925}}
\showeprint[arXiv]{2308.02925}


\bibitem[Wu et~al\mbox{.}(2019)]%
        {DBLP:conf/aaai/WuT0WXT19}
\bibfield{author}{\bibinfo{person}{Shu Wu}, \bibinfo{person}{Yuyuan Tang}, \bibinfo{person}{Yanqiao Zhu}, \bibinfo{person}{Liang Wang}, \bibinfo{person}{Xing Xie}, {and} \bibinfo{person}{Tieniu Tan}.} \bibinfo{year}{2019}\natexlab{}.
\newblock \showarticletitle{Session-Based Recommendation with Graph Neural Networks}. In \bibinfo{booktitle}{\emph{The Thirty-Third {AAAI} Conference on Artificial Intelligence, {AAAI} 2019, The Thirty-First Innovative Applications of Artificial Intelligence Conference, {IAAI} 2019, The Ninth {AAAI} Symposium on Educational Advances in Artificial Intelligence, {EAAI} 2019, Honolulu, Hawaii, USA, January 27 - February 1, 2019}}. \bibinfo{publisher}{{AAAI} Press}, \bibinfo{pages}{346--353}.
\newblock
\href{https://doi.org/10.1609/AAAI.V33I01.3301346}{doi:\nolinkurl{10.1609/AAAI.V33I01.3301346}}


\bibitem[Xie et~al\mbox{.}(2022)]%
        {cl4srec}
\bibfield{author}{\bibinfo{person}{Xu Xie}, \bibinfo{person}{Fei Sun}, \bibinfo{person}{Zhaoyang Liu}, \bibinfo{person}{Shiwen Wu}, \bibinfo{person}{Jinyang Gao}, \bibinfo{person}{Jiandong Zhang}, \bibinfo{person}{Bolin Ding}, {and} \bibinfo{person}{Bin Cui}.} \bibinfo{year}{2022}\natexlab{}.
\newblock \showarticletitle{Contrastive Learning for Sequential Recommendation}. In \bibinfo{booktitle}{\emph{38th {IEEE} International Conference on Data Engineering, {ICDE} 2022, Kuala Lumpur, Malaysia, May 9-12, 2022}}. \bibinfo{publisher}{{IEEE}}, \bibinfo{pages}{1259--1273}.
\newblock
\href{https://doi.org/10.1109/ICDE53745.2022.00099}{doi:\nolinkurl{10.1109/ICDE53745.2022.00099}}


\bibitem[Xu et~al\mbox{.}(2024)]%
        {tedrec}
\bibfield{author}{\bibinfo{person}{Lanling Xu}, \bibinfo{person}{Zhen Tian}, \bibinfo{person}{Bingqian Li}, \bibinfo{person}{Junjie Zhang}, \bibinfo{person}{Daoyuan Wang}, \bibinfo{person}{Hongyu Wang}, \bibinfo{person}{Jinpeng Wang}, \bibinfo{person}{Sheng Chen}, {and} \bibinfo{person}{Wayne~Xin Zhao}.} \bibinfo{year}{2024}\natexlab{}.
\newblock \showarticletitle{Sequence-level Semantic Representation Fusion for Recommender Systems}. In \bibinfo{booktitle}{\emph{Proceedings of the 33rd {ACM} International Conference on Information and Knowledge Management, {CIKM} 2024, Boise, ID, USA, October 21-25, 2024}}, \bibfield{editor}{\bibinfo{person}{Edoardo Serra} {and} \bibinfo{person}{Francesca Spezzano}} (Eds.). \bibinfo{publisher}{{ACM}}, \bibinfo{pages}{5015--5022}.
\newblock
\href{https://doi.org/10.1145/3627673.3680037}{doi:\nolinkurl{10.1145/3627673.3680037}}


\bibitem[Xu et~al\mbox{.}(2025b)]%
        {astro}
\bibfield{author}{\bibinfo{person}{Songpei Xu}, \bibinfo{person}{Shijia Wang}, \bibinfo{person}{Da Guo}, \bibinfo{person}{Xianwen Guo}, \bibinfo{person}{Qiang Xiao}, \bibinfo{person}{Fangjian Li}, {and} \bibinfo{person}{Chuanjiang Luo}.} \bibinfo{year}{2025}\natexlab{b}.
\newblock \showarticletitle{An Efficient Large Recommendation Model: Towards a Resource-Optimal Scaling Law}.
\newblock \bibinfo{journal}{\emph{CoRR}}  \bibinfo{volume}{abs/2502.09888} (\bibinfo{year}{2025}).
\newblock
\href{https://doi.org/10.48550/ARXIV.2502.09888}{doi:\nolinkurl{10.48550/ARXIV.2502.09888}}
\showeprint[arXiv]{2502.09888}


\bibitem[Xu et~al\mbox{.}(2025a)]%
        {softcot}
\bibfield{author}{\bibinfo{person}{Yige Xu}, \bibinfo{person}{Xu Guo}, \bibinfo{person}{Zhiwei Zeng}, {and} \bibinfo{person}{Chunyan Miao}.} \bibinfo{year}{2025}\natexlab{a}.
\newblock \bibinfo{title}{SoftCoT: Soft Chain-of-Thought for Efficient Reasoning with LLMs}.
\newblock
\showeprint[arxiv]{2502.12134}~[cs.CL]
\urldef\tempurl%
\url{https://arxiv.org/abs/2502.12134}
\showURL{%
\tempurl}


\bibitem[Yan et~al\mbox{.}(2019)]%
        {cosrec}
\bibfield{author}{\bibinfo{person}{An Yan}, \bibinfo{person}{Shuo Cheng}, \bibinfo{person}{Wang{-}Cheng Kang}, \bibinfo{person}{Mengting Wan}, {and} \bibinfo{person}{Julian~J. McAuley}.} \bibinfo{year}{2019}\natexlab{}.
\newblock \showarticletitle{CosRec: 2D Convolutional Neural Networks for Sequential Recommendation}. In \bibinfo{booktitle}{\emph{Proceedings of the 28th {ACM} International Conference on Information and Knowledge Management, {CIKM} 2019, Beijing, China, November 3-7, 2019}}, \bibfield{editor}{\bibinfo{person}{Wenwu Zhu}, \bibinfo{person}{Dacheng Tao}, \bibinfo{person}{Xueqi Cheng}, \bibinfo{person}{Peng Cui}, \bibinfo{person}{Elke~A. Rundensteiner}, \bibinfo{person}{David Carmel}, \bibinfo{person}{Qi~He}, {and} \bibinfo{person}{Jeffrey~Xu Yu}} (Eds.). \bibinfo{publisher}{{ACM}}, \bibinfo{pages}{2173--2176}.
\newblock
\href{https://doi.org/10.1145/3357384.3358113}{doi:\nolinkurl{10.1145/3357384.3358113}}


\bibitem[Yan et~al\mbox{.}(2025)]%
        {lum}
\bibfield{author}{\bibinfo{person}{Bencheng Yan}, \bibinfo{person}{Shilei Liu}, \bibinfo{person}{Zhiyuan Zeng}, \bibinfo{person}{Zihao Wang}, \bibinfo{person}{Yizhen Zhang}, \bibinfo{person}{Yujin Yuan}, \bibinfo{person}{Langming Liu}, \bibinfo{person}{Jiaqi Liu}, \bibinfo{person}{Di Wang}, \bibinfo{person}{Wenbo Su}, \bibinfo{person}{Pengjie Wang}, \bibinfo{person}{Jian Xu}, {and} \bibinfo{person}{Bo Zheng}.} \bibinfo{year}{2025}\natexlab{}.
\newblock \showarticletitle{Unlocking Scaling Law in Industrial Recommendation Systems with a Three-step Paradigm based Large User Model}.
\newblock \bibinfo{journal}{\emph{CoRR}}  \bibinfo{volume}{abs/2502.08309} (\bibinfo{year}{2025}).
\newblock
\href{https://doi.org/10.48550/ARXIV.2502.08309}{doi:\nolinkurl{10.48550/ARXIV.2502.08309}}
\showeprint[arXiv]{2502.08309}


\bibitem[Yang et~al\mbox{.}(2025)]%
        {reasonflux}
\bibfield{author}{\bibinfo{person}{Ling Yang}, \bibinfo{person}{Zhaochen Yu}, \bibinfo{person}{Bin Cui}, {and} \bibinfo{person}{Mengdi Wang}.} \bibinfo{year}{2025}\natexlab{}.
\newblock \showarticletitle{ReasonFlux: Hierarchical {LLM} Reasoning via Scaling Thought Templates}.
\newblock \bibinfo{journal}{\emph{CoRR}}  \bibinfo{volume}{abs/2502.06772} (\bibinfo{year}{2025}).
\newblock
\href{https://doi.org/10.48550/ARXIV.2502.06772}{doi:\nolinkurl{10.48550/ARXIV.2502.06772}}
\showeprint[arXiv]{2502.06772}


\bibitem[Yang et~al\mbox{.}(2024)]%
        {bot}
\bibfield{author}{\bibinfo{person}{Ling Yang}, \bibinfo{person}{Zhaochen Yu}, \bibinfo{person}{Tianjun Zhang}, \bibinfo{person}{Shiyi Cao}, \bibinfo{person}{Minkai Xu}, \bibinfo{person}{Wentao Zhang}, \bibinfo{person}{Joseph~E. Gonzalez}, {and} \bibinfo{person}{Bin Cui}.} \bibinfo{year}{2024}\natexlab{}.
\newblock \showarticletitle{Buffer of Thoughts: Thought-Augmented Reasoning with Large Language Models}. In \bibinfo{booktitle}{\emph{Advances in Neural Information Processing Systems 38: Annual Conference on Neural Information Processing Systems 2024, NeurIPS 2024, Vancouver, BC, Canada, December 10 - 15, 2024}}, \bibfield{editor}{\bibinfo{person}{Amir Globersons}, \bibinfo{person}{Lester Mackey}, \bibinfo{person}{Danielle Belgrave}, \bibinfo{person}{Angela Fan}, \bibinfo{person}{Ulrich Paquet}, \bibinfo{person}{Jakub~M. Tomczak}, {and} \bibinfo{person}{Cheng Zhang}} (Eds.).
\newblock
\urldef\tempurl%
\url{http://papers.nips.cc/paper\_files/paper/2024/hash/cde328b7bf6358f5ebb91fe9c539745e-Abstract-Conference.html}
\showURL{%
\tempurl}


\bibitem[Yuan et~al\mbox{.}(2019)]%
        {nextitnet}
\bibfield{author}{\bibinfo{person}{Fajie Yuan}, \bibinfo{person}{Alexandros Karatzoglou}, \bibinfo{person}{Ioannis Arapakis}, \bibinfo{person}{Joemon~M. Jose}, {and} \bibinfo{person}{Xiangnan He}.} \bibinfo{year}{2019}\natexlab{}.
\newblock \showarticletitle{A Simple Convolutional Generative Network for Next Item Recommendation}. In \bibinfo{booktitle}{\emph{Proceedings of the Twelfth {ACM} International Conference on Web Search and Data Mining, {WSDM} 2019, Melbourne, VIC, Australia, February 11-15, 2019}}, \bibfield{editor}{\bibinfo{person}{J.~Shane Culpepper}, \bibinfo{person}{Alistair Moffat}, \bibinfo{person}{Paul~N. Bennett}, {and} \bibinfo{person}{Kristina Lerman}} (Eds.). \bibinfo{publisher}{{ACM}}, \bibinfo{pages}{582--590}.
\newblock
\href{https://doi.org/10.1145/3289600.3290975}{doi:\nolinkurl{10.1145/3289600.3290975}}


\bibitem[Zhai et~al\mbox{.}(2024)]%
        {hstu}
\bibfield{author}{\bibinfo{person}{Jiaqi Zhai}, \bibinfo{person}{Lucy Liao}, \bibinfo{person}{Xing Liu}, \bibinfo{person}{Yueming Wang}, \bibinfo{person}{Rui Li}, \bibinfo{person}{Xuan Cao}, \bibinfo{person}{Leon Gao}, \bibinfo{person}{Zhaojie Gong}, \bibinfo{person}{Fangda Gu}, \bibinfo{person}{Jiayuan He}, \bibinfo{person}{Yinghai Lu}, {and} \bibinfo{person}{Yu Shi}.} \bibinfo{year}{2024}\natexlab{}.
\newblock \showarticletitle{Actions Speak Louder than Words: Trillion-Parameter Sequential Transducers for Generative Recommendations}. In \bibinfo{booktitle}{\emph{Forty-first International Conference on Machine Learning, {ICML} 2024, Vienna, Austria, July 21-27, 2024}}. \bibinfo{publisher}{OpenReview.net}.
\newblock
\urldef\tempurl%
\url{https://openreview.net/forum?id=xye7iNsgXn}
\showURL{%
\tempurl}


\bibitem[Zhang et~al\mbox{.}(2024)]%
        {scale-law4sr}
\bibfield{author}{\bibinfo{person}{Gaowei Zhang}, \bibinfo{person}{Yupeng Hou}, \bibinfo{person}{Hongyu Lu}, \bibinfo{person}{Yu Chen}, \bibinfo{person}{Wayne~Xin Zhao}, {and} \bibinfo{person}{Ji{-}Rong Wen}.} \bibinfo{year}{2024}\natexlab{}.
\newblock \showarticletitle{Scaling Law of Large Sequential Recommendation Models}. In \bibinfo{booktitle}{\emph{Proceedings of the 18th {ACM} Conference on Recommender Systems, RecSys 2024, Bari, Italy, October 14-18, 2024}}, \bibfield{editor}{\bibinfo{person}{Tommaso~Di Noia}, \bibinfo{person}{Pasquale Lops}, \bibinfo{person}{Thorsten Joachims}, \bibinfo{person}{Katrien Verbert}, \bibinfo{person}{Pablo Castells}, \bibinfo{person}{Zhenhua Dong}, {and} \bibinfo{person}{Ben London}} (Eds.). \bibinfo{publisher}{{ACM}}, \bibinfo{pages}{444--453}.
\newblock
\href{https://doi.org/10.1145/3640457.3688129}{doi:\nolinkurl{10.1145/3640457.3688129}}


\bibitem[Zhang et~al\mbox{.}(2025)]%
        {stream-rec}
\bibfield{author}{\bibinfo{person}{Junjie Zhang}, \bibinfo{person}{Beichen Zhang}, \bibinfo{person}{Wenqi Sun}, \bibinfo{person}{Hongyu Lu}, \bibinfo{person}{Wayne~Xin Zhao}, \bibinfo{person}{Yu Chen}, {and} \bibinfo{person}{Ji{-}Rong Wen}.} \bibinfo{year}{2025}\natexlab{}.
\newblock \showarticletitle{Slow Thinking for Sequential Recommendation}.
\newblock \bibinfo{journal}{\emph{CoRR}}  \bibinfo{volume}{abs/2504.09627} (\bibinfo{year}{2025}).
\newblock
\href{https://doi.org/10.48550/ARXIV.2504.09627}{doi:\nolinkurl{10.48550/ARXIV.2504.09627}}
\showeprint[arXiv]{2504.09627}


\bibitem[Zhao et~al\mbox{.}(2022)]%
        {recbole2.0}
\bibfield{author}{\bibinfo{person}{Wayne~Xin Zhao}, \bibinfo{person}{Yupeng Hou}, \bibinfo{person}{Xingyu Pan}, \bibinfo{person}{Chen Yang}, \bibinfo{person}{Zeyu Zhang}, \bibinfo{person}{Zihan Lin}, \bibinfo{person}{Jingsen Zhang}, \bibinfo{person}{Shuqing Bian}, \bibinfo{person}{Jiakai Tang}, \bibinfo{person}{Wenqi Sun}, \bibinfo{person}{Yushuo Chen}, \bibinfo{person}{Lanling Xu}, \bibinfo{person}{Gaowei Zhang}, \bibinfo{person}{Zhen Tian}, \bibinfo{person}{Changxin Tian}, \bibinfo{person}{Shanlei Mu}, \bibinfo{person}{Xinyan Fan}, \bibinfo{person}{Xu Chen}, {and} \bibinfo{person}{Ji{-}Rong Wen}.} \bibinfo{year}{2022}\natexlab{}.
\newblock \showarticletitle{RecBole 2.0: Towards a More Up-to-Date Recommendation Library}. In \bibinfo{booktitle}{\emph{Proceedings of the 31st {ACM} International Conference on Information {\&} Knowledge Management, Atlanta, GA, USA, October 17-21, 2022}}, \bibfield{editor}{\bibinfo{person}{Mohammad~Al Hasan} {and} \bibinfo{person}{Li~Xiong}} (Eds.). \bibinfo{publisher}{{ACM}}, \bibinfo{pages}{4722--4726}.
\newblock
\href{https://doi.org/10.1145/3511808.3557680}{doi:\nolinkurl{10.1145/3511808.3557680}}


\bibitem[Zhao et~al\mbox{.}(2021)]%
        {recbole}
\bibfield{author}{\bibinfo{person}{Wayne~Xin Zhao}, \bibinfo{person}{Shanlei Mu}, \bibinfo{person}{Yupeng Hou}, \bibinfo{person}{Zihan Lin}, \bibinfo{person}{Yushuo Chen}, \bibinfo{person}{Xingyu Pan}, \bibinfo{person}{Kaiyuan Li}, \bibinfo{person}{Yujie Lu}, \bibinfo{person}{Hui Wang}, \bibinfo{person}{Changxin Tian}, \bibinfo{person}{Yingqian Min}, \bibinfo{person}{Zhichao Feng}, \bibinfo{person}{Xinyan Fan}, \bibinfo{person}{Xu Chen}, \bibinfo{person}{Pengfei Wang}, \bibinfo{person}{Wendi Ji}, \bibinfo{person}{Yaliang Li}, \bibinfo{person}{Xiaoling Wang}, {and} \bibinfo{person}{Ji{-}Rong Wen}.} \bibinfo{year}{2021}\natexlab{}.
\newblock \showarticletitle{RecBole: Towards a Unified, Comprehensive and Efficient Framework for Recommendation Algorithms}. In \bibinfo{booktitle}{\emph{{CIKM} '21: The 30th {ACM} International Conference on Information and Knowledge Management, Virtual Event, Queensland, Australia, November 1 - 5, 2021}}. \bibinfo{publisher}{{ACM}}, \bibinfo{pages}{4653--4664}.
\newblock


\bibitem[Zheng et~al\mbox{.}(2024)]%
        {lsvcr}
\bibfield{author}{\bibinfo{person}{Bowen Zheng}, \bibinfo{person}{Zihan Lin}, \bibinfo{person}{Enze Liu}, \bibinfo{person}{Chen Yang}, \bibinfo{person}{Enyang Bai}, \bibinfo{person}{Cheng Ling}, \bibinfo{person}{Wayne~Xin Zhao}, {and} \bibinfo{person}{Ji{-}Rong Wen}.} \bibinfo{year}{2024}\natexlab{}.
\newblock \showarticletitle{A Large Language Model Enhanced Sequential Recommender for Joint Video and Comment Recommendation}.
\newblock \bibinfo{journal}{\emph{CoRR}}  \bibinfo{volume}{abs/2403.13574} (\bibinfo{year}{2024}).
\newblock
\href{https://doi.org/10.48550/ARXIV.2403.13574}{doi:\nolinkurl{10.48550/ARXIV.2403.13574}}
\showeprint[arXiv]{2403.13574}


\bibitem[Zhou et~al\mbox{.}(2020)]%
        {s3rec}
\bibfield{author}{\bibinfo{person}{Kun Zhou}, \bibinfo{person}{Hui Wang}, \bibinfo{person}{Wayne~Xin Zhao}, \bibinfo{person}{Yutao Zhu}, \bibinfo{person}{Sirui Wang}, \bibinfo{person}{Fuzheng Zhang}, \bibinfo{person}{Zhongyuan Wang}, {and} \bibinfo{person}{Ji{-}Rong Wen}.} \bibinfo{year}{2020}\natexlab{}.
\newblock \showarticletitle{S3-Rec: Self-Supervised Learning for Sequential Recommendation with Mutual Information Maximization}. In \bibinfo{booktitle}{\emph{{CIKM} '20: The 29th {ACM} International Conference on Information and Knowledge Management, Virtual Event, Ireland, October 19-23, 2020}}, \bibfield{editor}{\bibinfo{person}{Mathieu d'Aquin}, \bibinfo{person}{Stefan Dietze}, \bibinfo{person}{Claudia Hauff}, \bibinfo{person}{Edward Curry}, {and} \bibinfo{person}{Philippe Cudr{\'{e}}{-}Mauroux}} (Eds.). \bibinfo{publisher}{{ACM}}, \bibinfo{pages}{1893--1902}.
\newblock
\href{https://doi.org/10.1145/3340531.3411954}{doi:\nolinkurl{10.1145/3340531.3411954}}


\bibitem[Zhou et~al\mbox{.}(2022)]%
        {fmlp-rec}
\bibfield{author}{\bibinfo{person}{Kun Zhou}, \bibinfo{person}{Hui Yu}, \bibinfo{person}{Wayne~Xin Zhao}, {and} \bibinfo{person}{Ji{-}Rong Wen}.} \bibinfo{year}{2022}\natexlab{}.
\newblock \showarticletitle{Filter-enhanced {MLP} is All You Need for Sequential Recommendation}. In \bibinfo{booktitle}{\emph{{WWW} '22: The {ACM} Web Conference 2022, Virtual Event, Lyon, France, April 25 - 29, 2022}}, \bibfield{editor}{\bibinfo{person}{Fr{\'{e}}d{\'{e}}rique Laforest}, \bibinfo{person}{Rapha{\"{e}}l Troncy}, \bibinfo{person}{Elena Simperl}, \bibinfo{person}{Deepak Agarwal}, \bibinfo{person}{Aristides Gionis}, \bibinfo{person}{Ivan Herman}, {and} \bibinfo{person}{Lionel M{\'{e}}dini}} (Eds.). \bibinfo{publisher}{{ACM}}, \bibinfo{pages}{2388--2399}.
\newblock
\href{https://doi.org/10.1145/3485447.3512111}{doi:\nolinkurl{10.1145/3485447.3512111}}


\bibitem[Zhou et~al\mbox{.}(2023)]%
        {ac-tsr}
\bibfield{author}{\bibinfo{person}{Peilin Zhou}, \bibinfo{person}{Qichen Ye}, \bibinfo{person}{Yueqi Xie}, \bibinfo{person}{Jingqi Gao}, \bibinfo{person}{Shoujin Wang}, \bibinfo{person}{Jae~Boum Kim}, \bibinfo{person}{Chenyu You}, {and} \bibinfo{person}{Sunghun Kim}.} \bibinfo{year}{2023}\natexlab{}.
\newblock \showarticletitle{Attention Calibration for Transformer-based Sequential Recommendation}. In \bibinfo{booktitle}{\emph{Proceedings of the 32nd {ACM} International Conference on Information and Knowledge Management, {CIKM} 2023, Birmingham, United Kingdom, October 21-25, 2023}}, \bibfield{editor}{\bibinfo{person}{Ingo Frommholz}, \bibinfo{person}{Frank Hopfgartner}, \bibinfo{person}{Mark Lee}, \bibinfo{person}{Michael Oakes}, \bibinfo{person}{Mounia Lalmas}, \bibinfo{person}{Min Zhang}, {and} \bibinfo{person}{Rodrygo L.~T. Santos}} (Eds.). \bibinfo{publisher}{{ACM}}, \bibinfo{pages}{3595--3605}.
\newblock
\href{https://doi.org/10.1145/3583780.3614785}{doi:\nolinkurl{10.1145/3583780.3614785}}


\end{thebibliography}
